\documentclass[]{emulateapj}
\usepackage{epsfig}
\usepackage{amsmath}
\usepackage{natbib}
\usepackage{url,hyperref}

\usepackage{graphicx}
\usepackage{epstopdf}

\usepackage[toc,page]{appendix}

\defcitealias{apogee}{S16}

\begin{document}

\title{A Color-Locus Method for Mapping $R_V$ Using Ensembles of Stars}

\author{Albert Lee}
\affiliation{Harvard University}
\author{Gregory M. Green}
\affiliation{Stanford University}
\author{Edward F. Schlafly}
\affiliation{Lawrence Berkeley National Laboratory}
\author{Douglas P. Finkbeiner}
\affiliation{Harvard University}
\author{William Burgett}
\affiliation{Giant Magellan Telescope}
\author{Ken Chambers}
\affiliation{University of Hawaii}
\author{Heather Flewelling}
\affiliation{University of Hawaii}
\author{Klaus Hodapp}
\affiliation{University of Hawaii}
\author{Nick Kaiser}
\affiliation{University of Hawaii}
\author{Rolf-Peter Kudritzki}
\affiliation{University of Hawaii}
\author{Eugene Magnier}
\affiliation{University of Hawaii}
\author{Nigel Metcalfe}
\affiliation{Durham University}
\author{Richard Wainscoat}
\affiliation{University of Hawaii}
\author{Christopher Waters}
\affiliation{University of Hawaii}

\date{\today}

\begin{abstract}
	We present a simple but effective technique for measuring angular variation in $R_V$ across the sky. We divide stars from the Pan-STARRS1 catalog into Healpix pixels and determine the posterior distribution of reddening and $R_V$ for each pixel using two independent Monte Carlo methods. We find the two methods to be self-consistent in the limits where they are expected to perform similarly. We also find some agreement with high-precision photometric studies of $R_V$ in Perseus and Ophiuchus, as well as with a map of reddening near the Galactic plane based on stellar spectra from APOGEE. 
While current studies of $R_V$ are mostly limited to isolated clouds, we have developed a systematic method for comparing $R_V$ values for the majority of observable dust. This is a proof of concept for a more rigorous Galactic reddening map.
\end{abstract}

\maketitle

\section{Introduction}

Interstellar dust is an important part of the Milky Way. It is deeply intertwined with the formation and evolution of stars and plays a critical role in the physical and chemical processes of the interstellar medium (ISM). It is also one of the principal foregrounds for other objects of interest within and beyond the Milky Way.  Understanding dust and variations in its optical properties will be crucial steps toward understanding the structure of the Galaxy and for making accurate photometric corrections.

The composition of dust, however, remains mostly a mystery. For the better part of a century we have known that there must be a population of fine particles in our Galaxy in order to account for the selective absorption and scattering of bluer wavelengths in observed spectra \citep{trumpler}.  \citet{whittet} and others showed that the extinction curve of dust varies significantly along different sightlines through the Galaxy. \citet{fitz1986} and two subsequent papers by the same authors demonstrated that most extinction curves can be described by analytic expressions with as few as four parameters along typical lines of sight. \citet[hereafter CCM]{cardelli} then showed that the parameters are largely correlated and thus proposed that the main component of variation in extinction curves can be characterized by a single parameter $R_V$. $R_V$ is a prudent choice of parameter because $A(V)/E(B-V)$ is directly related to the slope of the extinction, which is the most salient differentiating feature for the optical and infrared wavelengths. There continues to be debate over whether $R_V$ captures all the significant reddening information for typical extinction curves. For example, \citet{fitz} and \citet{valencic} have further refined the single-parameter reddening law first proposed by CCM.  On the other hand, \citet{CardelliVar} and more recently \citet{fitzIR} have shown that one parameter may not be enough to reasonably describe all curves, especially as one ventures further into the ultraviolet or infrared. This suggests that we should avoid assuming any specific class of reddening laws and instead measure the actual per-wavelength reddening as empirically as possible.

So far, many groups have obtained different estimates of reddening in various regions of the sky using different techniques, and it is unclear how comparable these are.  New, large surveys like Pan-STARRS1 offer the opportunity to resolve these difficulties, by providing homogeneous photometry across a majority of the sky.  This motivates us to develop a technique to determine $R_V$ from photometry alone, allowing extinction curve measurements to be put on a common scale over most of the sky.  There have been several studies that use similar techniques. NICE, NICER \citep{NICER}, and NICEST \citep{NICEST} used photometry from the near-IR bands to estimate the total extinction along lines of sight. \citet{high} and \citet{SchlSDSS} used the visual bands to make similar estimates. Our study is different from the above examples in that we estimate the variation in the reddening in addition to the total extinction.

In order to make a full-sky map of reddening, we must have a population of well-characterized sources distributed throughout the whole sky for which we know the intrinsic spectrum. We can then compare the observed spectra to the same model intrinsic spectrum and estimate the selective attenuation - or reddening - of the sources. We must also have some means of evaluating the veracity of a map since there is no other full-sky map of reddening variation currently in existence. 

The PS1 catalog provides us with stars that satisfy these criteria. Although a single star measured in the five PS1 bands is not suitable for the pair method (unless the intrinsic colors of that star are known from some other technique, e.g., spectroscopy), a set of stars expected to be reddened by the same dust column has a well-understood locus in color-color space. We introduce two statistical methods for estimating reddening using ensembles of stars along different lines of sight. Stars are divided into Healpix pixels, with each pixel corresponding to a line of sight. The two methods were independently developed, allowing us to check for consistency.

The first method, which henceforth we refer to as the locus-shift method, makes the simplifying assumption that the majority of stars along a line of sight are behind the same dust column. This allows us to fit for the color excess of the entire locus of stars in a pixel, instead of fitting star by star. In the limit that this simple model is accurate, we gain a high signal-to-noise ratio of the position of the locus in color space with a greatly reduced computation time. 

The second method uses the Bayestar package \citep{greenMethods}. Given a set of stars in a pixel (i.e. a line of sight) and a model of stellar magnitudes based on their luminosity and metallicity, as well as priors on these parameters, Bayestar computes the full posterior of the possible realizations of dust reddening as a function of distance along the line of sight. While the publicly released map in \citet{Green3D} has $R_V$ fixed to $3.1$, in this study we let $R_V$ float. We expect this to provide the most precise map of reddening; however, it requires long computation times, and although we can use \citet[hereafter SFD]{SFD} to verify that it has measured extinction reliably, we have no way to determine its accuracy in estimating $R_V$. By comparing the locus-shift map with the Bayestar map, we can estimate some degree of confidence for each pixel in our maps.

In Section \ref{sec:ps1} we describe the data used to generate our maps. We explain the locus-shift method in Section \ref{sec:LS}, describing the locus model in Section \ref{sec:model}, and showing how we convert our reddening estimates to a single parameter $R_V$ in Section \ref{sec:rvfit}. In particular, we discuss the nuances of using a map of $R_V$ as a map of reddening, as well as the dangers of assuming that a simple reddening law holds in all cases.  In Section \ref{sec:mock} we discuss some of the insights into the locus-shift method gained from mock stars, and in Section \ref{sec:rvmap} we present our locus-shift reddening maps. We then explain how we adapted Bayestar as an alternate method for measuring $R_V$ in Section \ref{sec:bay}. In Section \ref{sec:analyze} we analyze our locus-shift results and compare them to those from Bayestar, as well as those from the independent study \citet[hereafter S16]{apogee}. Finally, in Section \ref{sec:conclude} we make concluding remarks.

\section{Pan-STARRS1}\label{sec:ps1}

This study uses stellar photometry from approximately $50$ million stars from the PS1 survey.
The PS1 survey \citep{PS1surveys} uses a wide-field telescope installed on the peak of Haleakala in Hawaii \citep{PS1_optics}. An array of $60$ $4800\times4800$ pixel CCDs is situated in the focal plane, and the system can swap between the $g$, $r$, $i$, $z$, $y$, and $w$ photometric filters \citep{PS1_optics, PS1_GPCB, PS1_GPCA}. A tunable laser was used to measure the filter transmission functions \citep{PS_lasercal}. The PS1 collaboration provides total expected transmission functions that account for the optical properties of the telescope. We use these to derive the attenuation in the PS1 bands as a function of reddening.

The survey itself covered the entire sky northward of declination $-30$, and the PS1 pipeline \citep{PS1_IPP, PS1pipeline, PS1sourcedet, PS1pixels2016} generated a catalog of most sources in this region brighter than roughly $22$ mag, depending on the band and data release \citep{PS1_photometry, PS1photast, PS1_astrometry, Magnier:2013}. The observations have been further calibrated and characterized by \citet{JTphoto} and \citet{Schlafly:2012}. We note that as the source detection algorithms improve, new iterations of the reduced data set have been released to the collaboration as Processing Versions \citep{PS1database}. This paper uses the latest Processing Version (PV3). In Sections \ref{sec:LS} and \ref{sec:bay}, we discuss our criteria for making a selection on this data set.

\begin{figure}
   \includegraphics[width=3.4in]{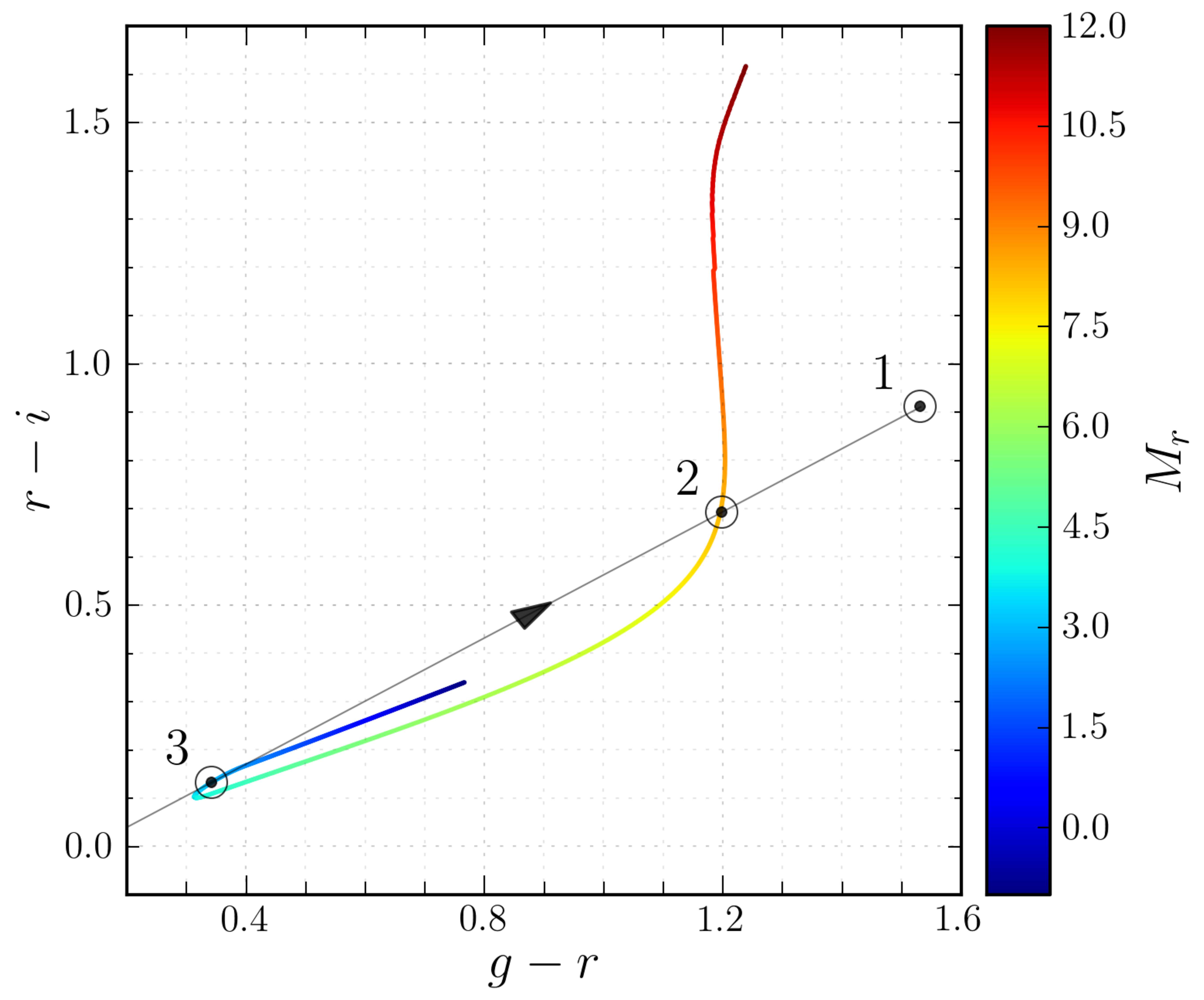}  
   \includegraphics[width=3.4in]{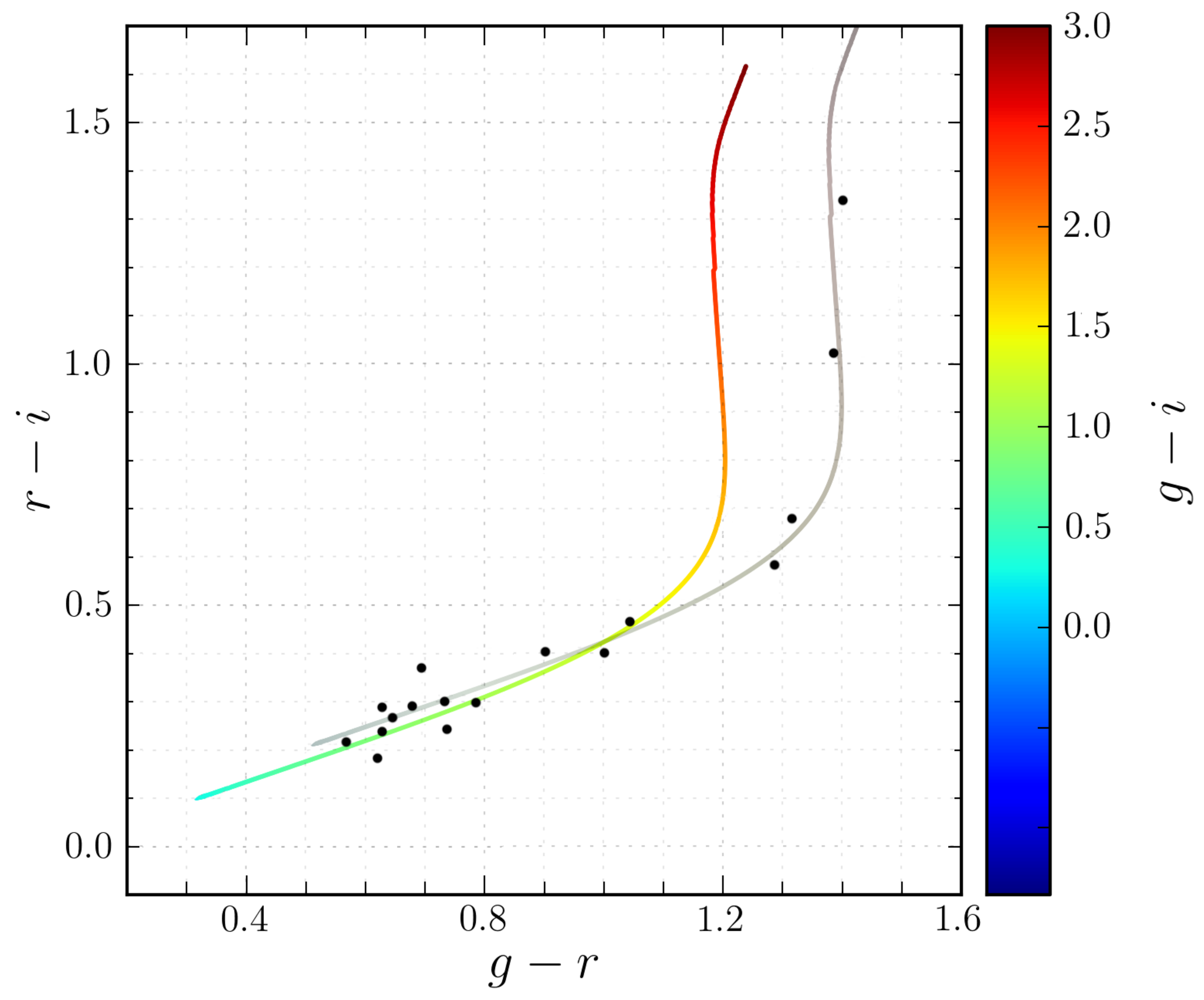}  
   \caption{Toy drawings illustrating the basic strategy for fitting for stellar distances and locus shifts. \\
	In the top panel, which we take directly from \citet{greenMethods}, we show the position $(1)$ of a source in color space superposed with a curve following our model for the main-sequence stellar locus. We move the source, purportedly a main-sequence star, backward in the opposite direction of the reddening vector until it intersects with the stellar locus at two possible photometric parallax solutions ($2$ and $3$). In this manner we can estimate the reddening of a star. \\
	In the bottom panel we repurpose Green's figure to show our model for the stellar locus (colored line) superposed with the set of stars along a line of sight (black circles). Assuming that all the stars are behind one cloud, we can model their positions in color space with a new reddened locus (gray line). We note that the locus-shift model does not include the bend at the blue end of the locus because it is fit to the observed distribution of stellar colors and does not incorporate information about stellar ages.\\}
\label{fig:toyfit}
\end{figure}

\section{The Reddening Law via the Locus-shift Method}\label{sec:LS}

We estimate the reddening in each pixel using a two-step process that fits an entire locus of stars simultaneously. For the first step, the locus shift, we take an unreddened locus model and fit for the shifts in color-color space that best approximate the position of the observed locus of stars from each pixel. Since PS1 has magnitudes in five bands and four possible informative colors, this is a four-parameter fit. For the second step, the reddening fit, we find the most likely $R_V$ and $E(B-V)$ based on the locus shifts. Since estimates of $R_V$ can be very different for different reddening laws, the $R_V$ results are only intended to be used to reveal very general properties of the ISM. We strongly encourage detailed comparisons to be made not with our $R_V$ values, but with the underlying color excesses. 
	
We have organized the subsections as follows. In Section \ref{sec:LSdef} we establish the general framework for the first step of our analysis, the locus shift. We provide a top-level description of our method for fitting shifts in color space and then proceed to lay out the details of the theory behind the method. Next, in Section \ref{sec:model}, we describe our procedure for empirically generating the most important component of our method, the locus model. We explain the limits of the model, then provide the technical details, and finally argue for the cuts used to generate the model. In Section \ref{sec:rvfit} we go over the second step of our analysis, the reddening fit. We show three different methods for converting locus-shift color excesses to reddenings (i.e. two numbers $E(B-V)$ and $R_V$). Finally, in Section \ref{sec:choose} we reiterate the importance of considering all dependencies when choosing a reddening model, and we argue for the particular choices made in this paper.

\subsection{Estimating Locus Shifts in Color Space}\label{sec:LSdef}

Please see the bottom panel of Figure \ref{fig:toyfit} for a schematic drawing of the locus fits described below. The locus-shift method has the advantage of being able to estimate the reddening of each color by combining information for multiple stars. 
The locus-shift method works best when there is a single predominant and relatively nearby source of reddening for a line of sight. In such a case most of the stars in the pixel are shifted in color space by the same reddening vector, resulting in the entire unreddened locus being shifted to its observed reddened position in color space. Fortunately, this situation is common for nearby dust clouds situated off the Galactic plane. In this limit we expect our results to agree well with other independent measurements of $R_V$, as well as the alternate Bayestar method we outline later in Section \ref{sec:bay}.

More rigorously, let $\vec{m_i}$ and $\vec{c_i}$ be the magnitudes and colors, respectively, of the $i^{ \text{th}}$ star in a Healpix pixel. We choose to subtract adjacent bands to obtain our colors, so that all color vectors $\vec{c}$ are in the $\left\{ g-r,r-i,i-z,z-y \right\}$ color space. The empirically determined unreddened locus $L_0$, i.e. the average stellar locus expected for lines of sight with no dust, is described by a curve through 
color-color space, $\vec{u_0}(x)$, as well as a distribution along the locus $p_L(x)$.  We find the $g-i$ color to be a reasonable choice for the parameter $x$, in that $\vec{u_0}(x)$ and $p_L(x)$ can both be accurately described by low-order polynomials.  We explain the derivation of the locus in detail in Section \ref{sec:locus}. Given some color excess $\vec{v}$, we get the reddened locus
\begin{equation}
\vec{u}(x,\vec{v}) = \vec{v} + \vec{u_0}(x).
\label{eq:vec_u}
\end{equation}
Since this is akin to moving the unreddened locus model in color space, in the context of our Monte Carlo model we will call $\vec{v}$ the locus shift and conceptually separate it from the color excess of a single source.

\begin{figure}
   \includegraphics[width=3.4in]{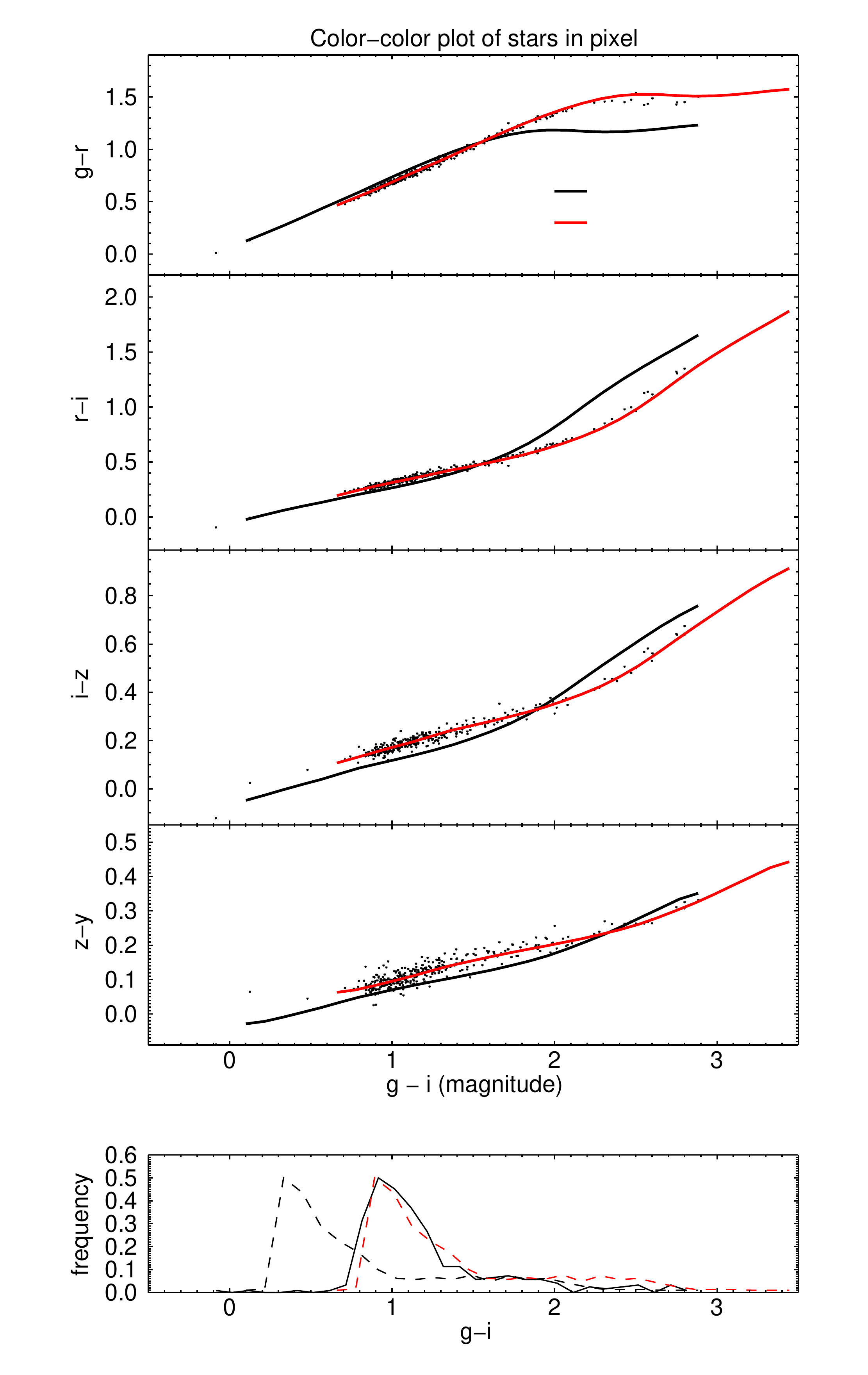}
   \caption{Example of a locus-shift fit. The Monte Carlo finds the shift in color space that brings the unreddened locus (solid black line) to a new position (solid red line) as close as possible to the set of mostly main-sequence stars (black dots) in a Healpix pixel. In the bottom panel we show the model's unreddened locus number density (dashed black line) and shifted number density (dashed red line) superposed with a histogram of the same set of stars.\\ }
\label{fig:loc}
\end{figure}

Figure \ref{fig:loc} shows an example of a locus-shift fit. The black line corresponds to the unreddened locus $\vec{u_0}$, and the red line is the reddened locus $\vec{u}$ after being displaced by $\vec{v}$. As expected, $\vec{u}$ is centered among the stellar colors $\vec{c_i}$ for this pixel.

For some position $x$ on the locus, the probability of a star's colors is given by a modified normal distribution in color excess
\begin{eqnarray}
\small{p_i(c_i | \vec{v},x)} &=& \tiny{\exp^\prime}\left[ \frac{\small{-(\vec{c_i}-\vec{u}(\vec{v},x))^T 
                                                          \Sigma_c^{-1} (\vec{c_i}-\vec{u}(\vec{v},x))}} {2} \right], \qquad  
\label{eq:p_star}\\
\exp^\prime(y) &=& A \tanh(\exp(y)/A), 
\end{eqnarray}
where $\Sigma_c$ is the covariance matrix derived from the uncertainties in stellar colors. If we let $\Sigma_c^{\alpha,\beta}$ be the element of $\Sigma_c$ corresponding to the colors $\alpha$ and $\beta$, and if we let $\Sigma_m^{\alpha,\beta}$ be the covariance between band magnitudes $\alpha$ and $\beta$, then given the basis of colors $\left\{ g-r,r-i,i-z,z-y \right\}$ and the basis of magnitudes $\left\{ g,r,i,z,y \right\}$, we get the relation
\begin{equation}
\Sigma_c^{\alpha,\beta} = \Sigma_m^{\alpha,\beta} - \Sigma_m^{\alpha,\beta+1} 
                            - \Sigma_m^{\alpha+1,\beta} + \Sigma_m^{\alpha+1,\beta+1},
\label{eq:sigmaC}
\end{equation}
as demonstrated in \citet{greenMethods}.  We choose $A$ so that $\exp^\prime(x)$ is a function that asymptotically approaches $\exp(x)$ for offsets less than $5\sigma$ and flattens out to a constant value when the total offset is beyond $5\sigma$.  These wings in the distribution allow us to reject outlier stars that deviate from the locus as a result of either not being on the main sequence or having drastically different reddening from the other stars in the pixel. Marginalizing over the locus, the likelihood for the colors of all the stars in a pixel is then
\begin{equation}
p(\left\{\vec{c}\right\} | \vec{v}) = \prod_i\int{dx{}\, p_i(c_i | \vec{v},x)\, p_L(x)}.
\label{eq:p_vec_c}
\end{equation}
Note that we have absorbed the variation in metallicity into the systematics and account for any corresponding uncertainties when we calculate $\Sigma_c$. We take the standard deviations of the residuals of each color from our model fits and add these to the diagonals of $\Sigma_c$. This of course accounts for any systematic, but we expect metallicity to be the largest component. The uncertainties are $0.021$, $0.021$, $0.020$, and $0.017$ for $g-r$, $r-i$, $i-z$, and $z-y$, respectively.

We fit for $\vec{v}$ using a Metropolis--Hastings Monte Carlo to sample from the posterior
\begin{equation} 
p(\vec{v} | \left\{\vec{c}\right\}) \propto p(\left\{\vec{c}\right\} | \vec{v}) p_v(\vec{v}).
\label{eq:p_pos}
\end{equation}
We generally use a flat prior $p_v(\vec{v})$. 
This allows us to apply empirically determined priors on $R_V$ later on in our analysis.

We save $100$ samples from each Monte Carlo chain of originally $1000$ samples. We discard the first $30$ saved samples as burn-in whenever we calculate any statistics. Each pixel has three chains with different initial conditions, giving us $210$ samples total. To test for convergence, we use a Gelman--Rubin diagnostic. 

We note that this technique is similar to stellar locus regression, which \citet{high} showed to be effective for making corrections for atmospheric and dust extinction when estimating the photometric redshifts of galaxies.

\subsection{The Locus Model}\label{sec:model}

Our stellar locus models are all derived empirically by fitting the colors of stars from regions of low dust emission in \citet{SFD}. This ensures that our fits are accurate in the low-dust limit, as we are not dependent on a theoretical model of the mass function and metallicity distribution that reconstructs observed stellar colors. 
However, our reliance on empirical stellar loci makes it challenging to know how accurate our stellar locus is in the Galactic plane, where no low reddening regions are available.  Moreover, low-latitude stars are more likely to be younger and more metal rich than typical high-latitude halo stars, leading us to expect our stellar locus to poorly represent stars directly in the Galactic plane.  We discuss ways to resolve this in Section \ref{sec:galvar} and also provide a mask to avoid problematic regions, as detailed throughout Section \ref{sec:rvmap}

\begin{figure}
   \leavevmode\epsfxsize=8.5cm\epsfbox{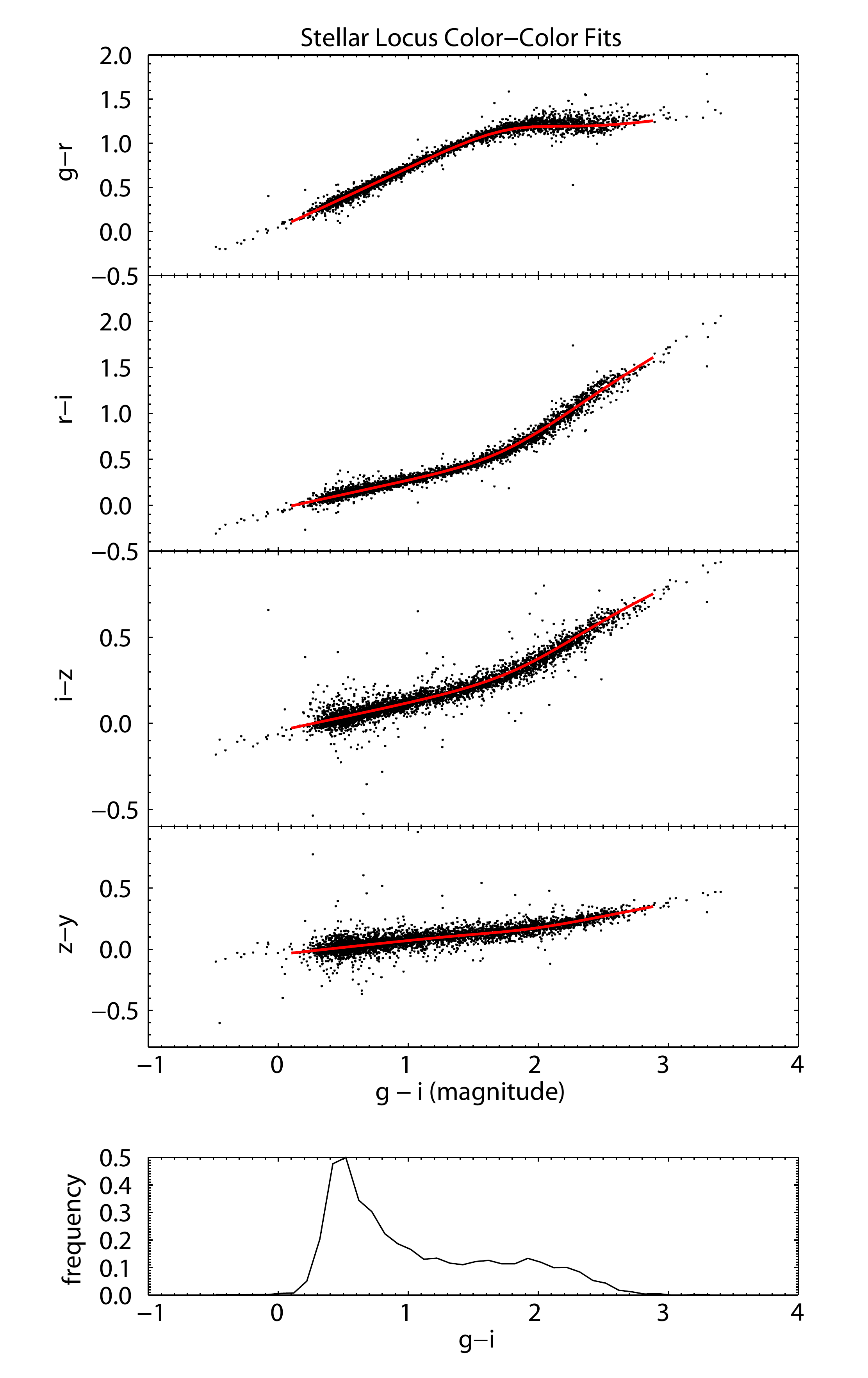} 
   \caption{Our stellar locus model superposed with a random subset of the stars against which the model was fit. As expected, the model is a perfect fit. The stars were selected from regions with $E(B-V)_{\text{SFD}}<0.012$ at a variety of Galactic latitudes, this particular example being from $|b|>30$. \\}
\label{fig:stellarLoc}
\end{figure}

\subsubsection{Empirical Stellar Loci}\label{sec:locus}

In order to make an accurate empirical model, we need a set of real stars completely unobscured by dust. However, since this is impossible, we correct the colors of stars by assuming a reddening law and using SFD \citep{SFD} as a template for dust optical depth. The locus model is then a best-fit curve to the unreddened stellar colors $\vec{c_u}$ for a set of stars in regions with an effective $E(B-V)_{\text{SFD}}<0.012$:
\begin{equation}
\vec{c_u}=\vec{c}_{E_{\text{SFD}}<0.012}-\vec{v_R}(R_V,E_{\text{SFD}}).
\end{equation}
We use the \citet{fitz} reddening law with $R_V$ set to $3.1$, as that has been found to be describe the mean reddening at low dust columns \citep{SchlSDSS}. Assuming that our locus model is correct, our choice of reddening law for this dereddening procedure has little effect 
on our locus-shift fits to real data for pixels with $E(B-V)_{\text{SFD}}>0.06$. Pixels with low dust do experience a bias, but since these regions inherently have little reddening information, they are masked in our results.

Figure \ref{fig:stellarLoc} shows an example unreddened locus model. It is a five-dimensional function of a single variable, the color $g-i$. Four of the dimensions describe the curve traced out by the mean locus through the space comprising the colors $g-r$, $r-i$, $i-z$, and $z-y$. The fifth dimension is the density of stars in $g-i$. The density effectively serves as a luminosity function, and we marginalize over this distribution when finding the likelihood of a star's colors. $g-i$ itself is a dummy variable for relating the five dimensions to one another, and it has no bearing on the result of the integral in the likelihood function in Equation \ref{eq:p_vec_c}.

\subsubsection{Cuts for Generating the Locus Model}\label{sec:LScuts}

To obtain the best loci for carrying out fits, we exclude stars that are not likely to be on the main sequence, as well as those with bad photometry. For the first case, we cut stars that lie blueward of our unreddened locus models. Since dust extinction can only move stars along the reddening vector in color-color space, any detection beyond the blue end of our locus is most likely a white dwarf or a quasar. We keep detections redder than our unreddened locus since obviously these are likely to be real stars obscured by dust. Stars that pass these cuts but are not in reality main-sequence stars should be effectively ignored in our fits owing to the wings in the probability distribution described in Eq. \ref{eq:p_star}.
For the second case, we cut stars that are fainter than $19$ mag in the $r$ band,
which excludes stars with low signal-to-noise ratios.

\subsection{Estimating the Reddening Law}\label{sec:rvfit}

Once we have a chain of locus shifts for each pixel, we find the $R_V$ and $E(B-V)$ that best approximate the distribution of $\vec{v}$.  This effectively entails a projection from the four-dimensional color space down into a two-dimensional reddening law subspace. We employ a few different methods depending on the comparison we wish to make. By decoupling $R_V$ from the Monte Carlo locus fits, we can quickly iterate over different reddening laws and save computation time for more accurate locus shifts. This lets us be flexible with the extinction models we use, instead of presupposing a specific reddening law. Since there have not been many comprehensive studies of $R_V$ variation, we believe that this is crucial for extracting the most accurate and useful information from our fits.

However, there are some dangers in arbitrarily converting locus shifts to $R_V$ for comparison with other data sets, especially when other photometric surveys may include different sets of stars and different passbands, so we recommend using the locus shifts in our data product as much as possible. If the reader still finds it necessary to obtain some $R_V$ value, we describe below three different methods for estimating reddenings and list their pros and cons, both to provide the reader with options and to inform them of the types of pitfalls we encountered. We suggest the proxy ${R_V}_{gy}$ defined below because it is most closely tied to the canonical definition of $R_V = A(V)/E(B-V)$,
but we emphasize that the link between $R_V$ and $E(B-V)$ is model dependent. For especially noisy data, the reader may be better served by the first method listed below, a gradient descent fit to a reddening law of choice, since it uses information from all four color shifts. Furthermore, we recommend extra precaution when converting between different formulations of $R_V$, as demonstrated in Figure \ref{fig:redlaw}

\subsubsection{Gradient Descent}\label{sec:gradient}

When we wish to determine the expected $R_V$ for a specific reddening law, we use a gradient descent algorithm with a chi-square goodness of fit.

Let $\vec{v_R}(R_V,E)$ be a reddening law function that returns a set of color shifts, given the parameters $R_V$ and reddening $E\equiv E(B-V)$ (this is the same numerically integrated function defined by Equation \ref{eq:redvec}). The likelihood of a color shift $\vec{v}$ is then
\begin{equation}
p_R(\vec{v}|R_V,E) = N\left( \vec{v}|\vec{v_R}(R_V,E),\Sigma_v \right),
\end{equation}
where $N(\vec{v}|\vec{\mu},\Sigma)$ is a multivariate normal function with mean $\vec{\mu}$ and covariance matrix $\Sigma$ evaluated at $\vec{v}$. $\Sigma_v$ is calculated directly from the distribution of samples in the locus-shift Monte Carlo chains.

The posterior is 
\begin{equation} 
p( R_V,E | \vec{v}) \propto p_R( \vec{v} | R_V,E ) p(R_V,E).
\label{eq:p_posRV}
\end{equation}
We find that the distribution of samples in $\vec{r} = (R_V,E)$ parameter space is roughly Gaussian. Therefore, we find it faster and sufficient to use a gradient descent search to determine the most likely $\vec{r}$. 

We fit a number of reddening laws to the locus shifts in order to find the most accurate one. For the objective function, we use the mean $\chi^2$ of all pixels with $0.3 < E(B-V) < 0.6$ since in this regime the fits should neither be sensitive to our choice of locus model nor to the unique reddening properties found in dense clouds. For CCM, \citet{Odon}, and \citet[hereafter F99]{fitz}, we get reduced $\chi^2$ values of $45.9$, $112.6$, and $11.5$, respectively. Therefore, we use F99 to generate the unreddened locus models above. However, the consistently high $\chi^2$ values show that none of the standard reddening models capture the full variability in reddening in the Milky Way.

A more principled method would be to convert the samples to $R_V$ samples using a function that maps $\vec{r}$ to $R_V$. Although this does not address the need for more than one reddening parameter, it allows us to examine the distribution of the posterior in $R_V$. In the next section we demonstrate how to make a linear projection that is a good approximation of such a mapping function.

\subsubsection{Linear Transforms}\label{sec:linear}

Since our fit results are coordinates in color space, and since the two-dimensional manifold mapped by $R_V$ and $E(B-V)$ through color space has very little curvature regardless of the reddening law we choose, we are able to linearize the conversion from locus shift to a coordinate in some affine $(R_V',E(B-V)')$ subspace, i.e. $(R_V',E(B-V)') = \vec{r'} = M\vec{v}$, for some projection $M$, without much loss of information. In fact, for some function given by a reddening law like F99, for example, $R_V=f_{F99}(\vec{v})$, we can reproduce the relation to within a few hundredths of $R_V$ using only a second-order polynomial of $\vec{r}'$.

This property of $\vec{r}$ and $\vec{v}$ in principle allows us to compare $R_V$ values derived by the locus-shift method with those derived using principal components in \citetalias{apogee}. It also allows us to quickly calculate a distribution of $R_V$. By taking the first principal component to be the true mean reddening vector for the Milky Way, we can express $\vec{r}$ as $c_0 \vec{p_0} + c_1 \vec{p_1}$ and get $R_V$ as a function of the ratio of $c_1/c_0$. Unfortunately, it turns out that the mean reddening vector given in \citetalias{apogee} and the one derived by this study using PS1 stars are not the same. We have identified three main contributing factors.  First, the \citetalias{apogee} study uses three Two Micron All Sky Survey bands and two \textit{WISE} bands in addition to the five PS1 bands used here.  This means that vectors that are orthogonal in the 10-band space will no longer be so in the five-band space.  Our different conventions for linearizing the reddening vector necessitate this.  This is largely not an issue because we can simply project one linearization to the other, and indeed doing so shows that they span roughly the same planes.

The second factor, which is related to the first, is the result of both studies being insensitive to any gray components $\vec{g}$ in the reddening vector.  This means that we must fit $\vec{g}$ to a reasonable reddening law in order to obtain color ratios.  Doing this fit, with five bands or $10$ bands, produces different mean reddening vectors.  
The 3D subspaces spanned by $\vec{p_0}$, $\vec{p_1}$, and $\vec{g}$ for both studies show even closer agreement with each other than the subspaces spanned by just $\vec{p_0}$ and $\vec{p_1}$.  However, we cannot simply force them to agree by fitting $\vec{g_{LS}}$ to $\vec{g_{S16}}$ to each other since both vectors are degenerate with the distance modulus and thus potentially encode real physical properties of the stars.

The third factor is that the APOGEE survey focused on giants, whereas PS1 only has a photometric limit, implying that most sources will be main-sequence stars. This means that, given some detection limit, the two surveys include stars at different distances and behind different dust columns. Additionally, stars with different spectra have different integrated fluxes through bandpasses and thus have different reddening vectors for the same extinction curve.  By comparing sightlines with high evidence of having only one predominant source of reddening, we can mitigate the former effect, but in general we expect the mean reddening vector to be different.

On the other hand, these effects are irrelevant when applying two different analyses to the same data set. Therefore, we use a principal component formulation of $R_V$ when comparing our locus-shift results with Bayestar results in Section \ref{sec:bayLS}. Projecting the \citetalias{apogee} principal components to be orthogonal in the PS1 bands, we get the vectors presented in the table below.
\begin{center}
  \begin{tabular}{ | l | c  c  c  c  c | }
    \hline
     & $g$ & $r$ & $i$ & $z$ & $y$ \\
		\hline
    $\vec{p_0}$  &  $0.390$  & $0.127$ & $-0.0620$ & $-0.185$  & $-0.269$ \\ 
    $\vec{p_1}$  &  $-0.100$ & $0.0882$ & $0.144$  &  $0.0164$ & $-0.148$ \\
    \hline
  \end{tabular}
\end{center}
We convert the linear transformations to reddening parameters via the formulae
\begin{eqnarray} 
R_V = 3.516 + 4.34 c_1/c_0 \label{eq:PCformula}, \\
E(B-V) = c_0 / 3.2,
\end{eqnarray}
which have been adapted from \citetalias{apogee} and fit to F99.

In summary, this means that our locus-shift samples are linearizable when comparing $R_V$ values derived from band magnitudes from similar wavelength ranges and similar sets of sources. However, when extrapolating to different wavelengths or using different sources, we must be careful to make the correct linear projection and to use a self-consistent definition of $R_V$. We suggest a method for addressing these issues in the next section.

\begin{figure}
	 \includegraphics[width=2.9in]{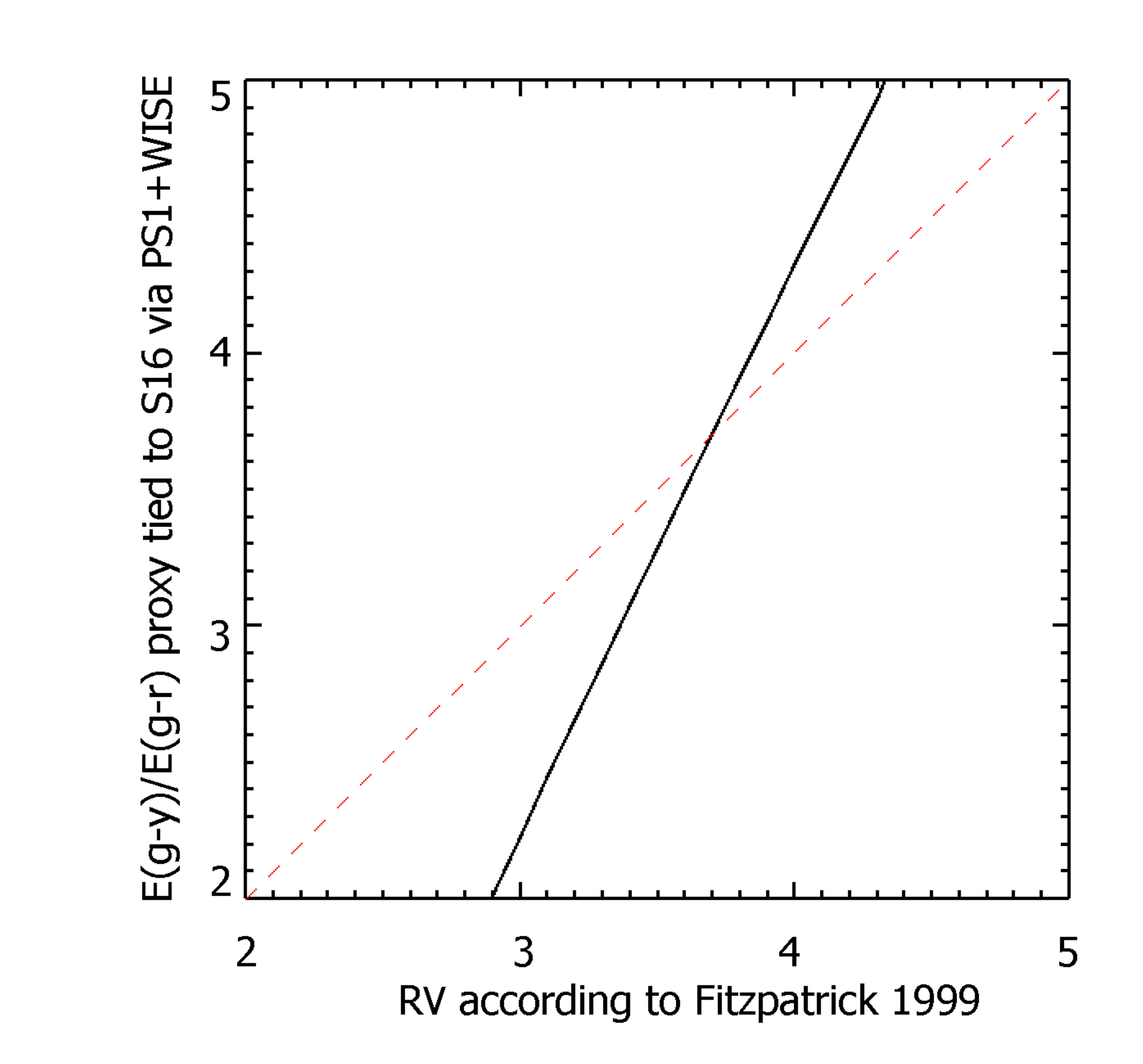} 
	 \includegraphics[width=2.9in]{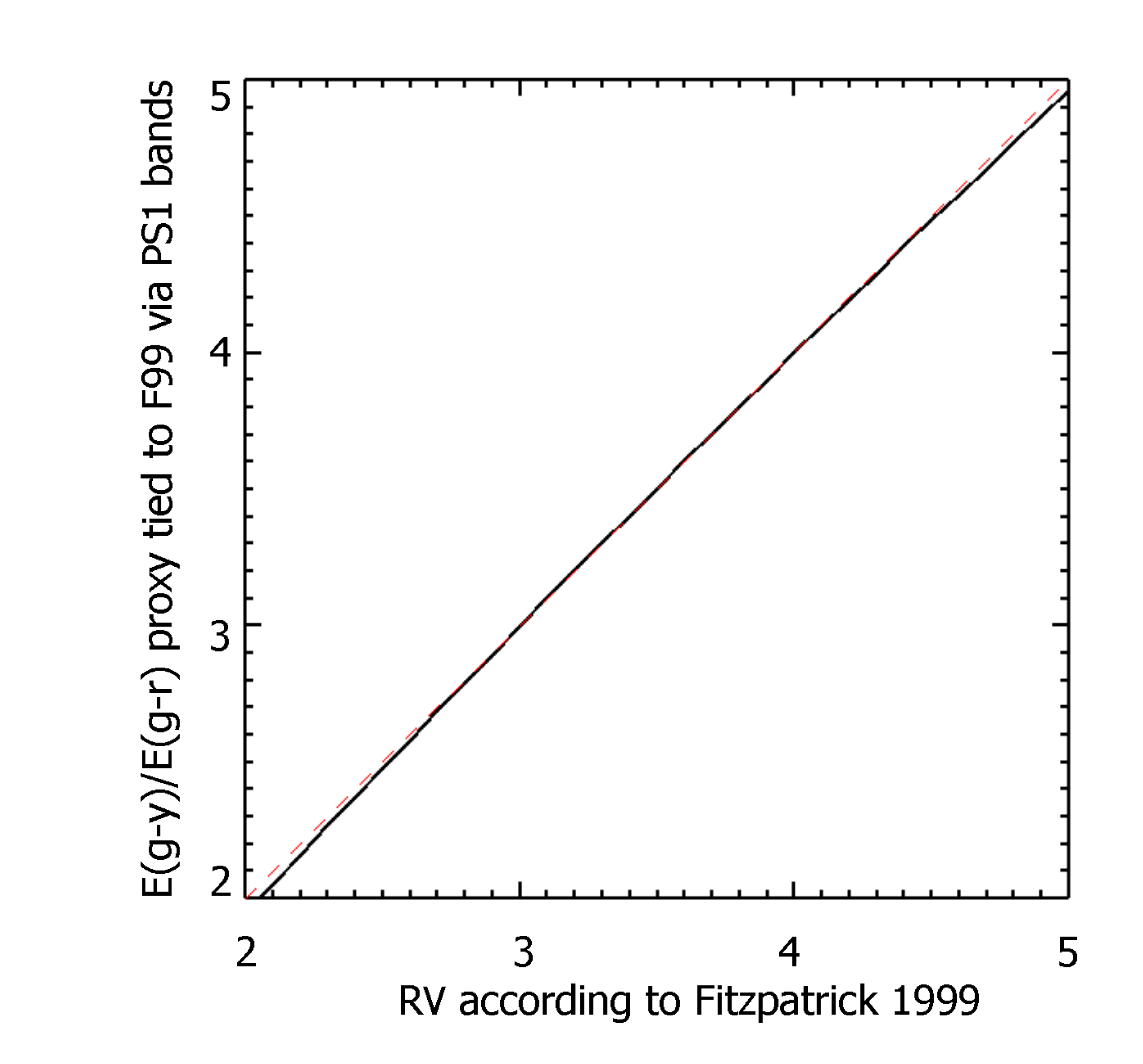}  
   \caption{These plots show how sensitive $R_V$ estimates are to the choice of parameterization. When converting a locus shift to some value of $R_V$ we must assume some set of color ratios in order to fix the undetermined gray component. Tying color ratios to a semitheoretical reddening law, such as F99, is one way to do this. The conversion must additionally be flexible enough to fit reasonably well with a variety of reddening scenarios and data sets while still being accurate and informative. This is difficult to accomplish because different surveys use different filters and observe different sets of stars. Since surveys generally measure the integrated flux through some filters, some information is lost. We suggest using the $R_V$ proxy defined by Equations \ref{eq:gygr} and \ref{eq:Schlafly} in Section \ref{sec:rvdef}, but even this has potential issues if used improperly.\\
	 For both panels we generate some locus shifts assuming that F99 is the true reddening in a dust cloud. Then we convert the locus shifts back to $R_V$ using a proxy and compare with the original $R_V$ according to F99. In the top panel we use the proxy $R_V = 4.6 E(g-y)/E(g-r)-8.2$, which is the formula we get if we substitute $y$ for $\text{W2}$ in Equation \ref{eq:Schlafly} using the relations given by $\vec{v}_{\text{S16}}$, the mean reddening vector in \citetalias{apogee}. In the bottom panel we use Equation \ref{eq:gygr}, whose constants have been obtained by fitting the proxy to F99 directly. As expected, the latter agrees more closely with F99 since we have optimized the constants to do so. The discrepancy in the top panel is not due to errors in either our models or those used in \citetalias{apogee}, but rather is the result of fitting the function to a model other than F99, i.e. $\vec{v}_{\text{S16}}$, which itself was accurate for the data in \citetalias{apogee}. In other words, although the F99 and \citetalias{apogee} extinction curves have similar behavior over a broad range of wavelengths, extrapolating them from a narrow range can lead to very different results.\\ 
	 We therefore recommend using the full locus-shift fits when making comparisons and only converting to $R_V$ when evaluating specific reddening laws such as F99.\\}
\label{fig:redlaw}
\end{figure}

\subsubsection{An $A(V)/E(B-V)$ Proxy}\label{sec:rvdef}

We are inclined to formulate a definition of $R_V$ that is as insensitive as possible to the above effects. The best solution would be to use a model of reddening that incorporates stellar types and dust extinction for a given set of photometric bands, and this will be the topic of future papers. In the meantime we adopt Equations \ref{eq:gygr} for reddenings from PS1 and Equation \ref{eq:Schlafly} for reddenings from \citetalias{apogee}. We choose them for some useful properties listed below, but note that there are many other reasonable formulations as well, especially due to the variability of the reddening vector:

\begin{equation}
{R_V}_{gy} = 2.2 \frac{E(g-y)}{E(g-r)} - 1.99, \\
\label{eq:gygr}
\end{equation}

\begin{equation}
{R_V}_{gW2} = 1.2 \frac{E(g-\text{W2})}{E(g-r)} - 1.18. \\
\label{eq:Schlafly}
\end{equation}

These definitions have the advantage of both being mathematically similar to the original formulation $R_V = A(V)/E(B-V)$ in that $E(g-r)$ behaves similarly to $E(B-V)$ and $E(g-y)$ and $E(g-\text{W2})$ are similar to $A_V$ for the sources involved. Equation \ref{eq:Schlafly}, as well as the strategy of using this functional form, was taken from \citetalias{apogee}. Equation \ref{eq:Schlafly} works specifically because \citetalias{apogee} matched data for sources from the PS1 and \textit{WISE} surveys, allowing them to correlate reddenings in the PS1 $g$, $r$, and \textit{WISE} 2 bands, and fit the relation to F99.  As long as the user is careful to use the correct linear coefficients, biases from selection effects should be minimized.
See Figure \ref{fig:redlaw} for details.

We also note that Equations \ref{eq:gygr} and \ref{eq:Schlafly} agree with each other for a wide range of $R_V$ values and typical reddenings because they have been fit to an independent empirical estimate of the typical reddening in the Galaxy (namely, F99). To be more precise, in as much as the true locus shift for a pixel lies at a point in color space close to the F99 color surface, both formulae will agree with each other and serve as excellent proxies for the canonical definition of $R_V$. Most other empirically derived reddening laws, such as CCM, lie somewhat close to F99 in color space, but not enough for perfect agreement.	This is especially true for the mean reddening vector provided by \citetalias{apogee}, and so we elect to tie our $R_V$ parameterization to F99 via Equation \ref{eq:gygr} when comparing to results from the principal component analysis in \citetalias{apogee}. 

We hope this further illustrates the need to use all the information in a locus shift or color excess when making reddening comparisons, rather than projecting down into $(R_V,E(B-V))$ space.

\subsection{Choosing an Extinction Model}\label{sec:choose}

To summarize, our reddening estimates first involve using a Metropolis--Hastings sampler to obtain the posterior distribution of locus shifts in color space.  Next, we convert the locus shifts to a reddening parameter $R_V$ according to some extinction model. We have described three different ways to make this conversion: a gradient descent fit to a reddening law, linearizing the color-excess information into reddening vectors, or calculating $R_V$ via a proxy formula tied to a model for the reddening in different bands.  For the range of wavelengths in the PS1 survey, all of these models are relatively linear, and for the majority of pixels covered by the survey, our locus-shift results tend to be very similar, or can be converted to one another with a simple linear transformation.  Since we are looking for correlations between our estimated $R_V$ values and those of other studies, in the context of this paper our results are robust for any choice of the three conversion methods.  

For all the plots in Section \ref{sec:rvmap} and onward we use the proxy definition of $R_V$ from Equation \ref{eq:gygr} (even though it does not affect our conclusions), since it is similar to the definition used in \citetalias{apogee} and because its physical motivation is straightforward.  This formula is tied to F99 because, as shown in Section \ref{sec:gradient}, F99 has the lowest $\chi^2$ of the standard reddening laws when fit to our locus-shift results. To estimate $E(B-V)$, we fit F99 to the locus shifts via gradient descent, as per Section \ref{sec:gradient}. The one exception to the above is where we use the principal component model from Section \ref{sec:linear} when comparing locus-shift fits to Bayestar fits owing to the latter having a natural linear formulation. If the reader needs to use some definition of $R_V$ to make their own comparison, they should carefully consider the pros and cons of each method listed above.

\section{Mock Stars}\label{sec:mock}

We test the reliability of the locus-shift method by running our analysis on a set of mock stars. We simulate the effects of dust on stellar colors in order to better  understand how our model fits should behave in various situations. The colors of stars are simulated by using the locus models as follows:
\begin{enumerate}
\item For each star, a point is chosen along the locus curve in color space.
\item The probability density of choosing a coordinate is specified by the $g-i$ distribution.
\item The color coordinates are displaced by a random color vector with a normal distribution. This accounts for the uncertainty in color measurements, as well as systematic uncertainties from  ignoring metallicity.
\item Next, the colors are shifted according to the reddening vectors corresponding to the dust column in front of each star.
\item Colors for a single star may be shifted several times to simulate multiple dust clouds with different reddening.
\end{enumerate}

The resulting set of reddened stellar colors allow us to test for a variety of cases. We generate mock stars for the following scenarios:
\begin{enumerate}
	\item All the stars are behind a single dust cloud. We simply shift all stellar colors in a pixel by the same reddening vector for this case.
	\item A fraction of the stars in a pixel have a different amount of reddening compared to others in the same pixel.  We expect this to occur when the dust column has spatial variation at subpixel angular scales, e.g. in dense cores and filaments.
	\item A fraction of the stars in a pixel are reddened by one more cloud in addition to a nearby cloud that reddens all the stars in the pixel.  This corresponds to the case where there are multiple layers of clouds along a line of sight.
\end{enumerate}

Sanity checks of the Bayestar method can be found in \citet{greenMethods} and \citet{Green3D}

\begin{figure}
   \includegraphics[height=3.3in,angle=90]{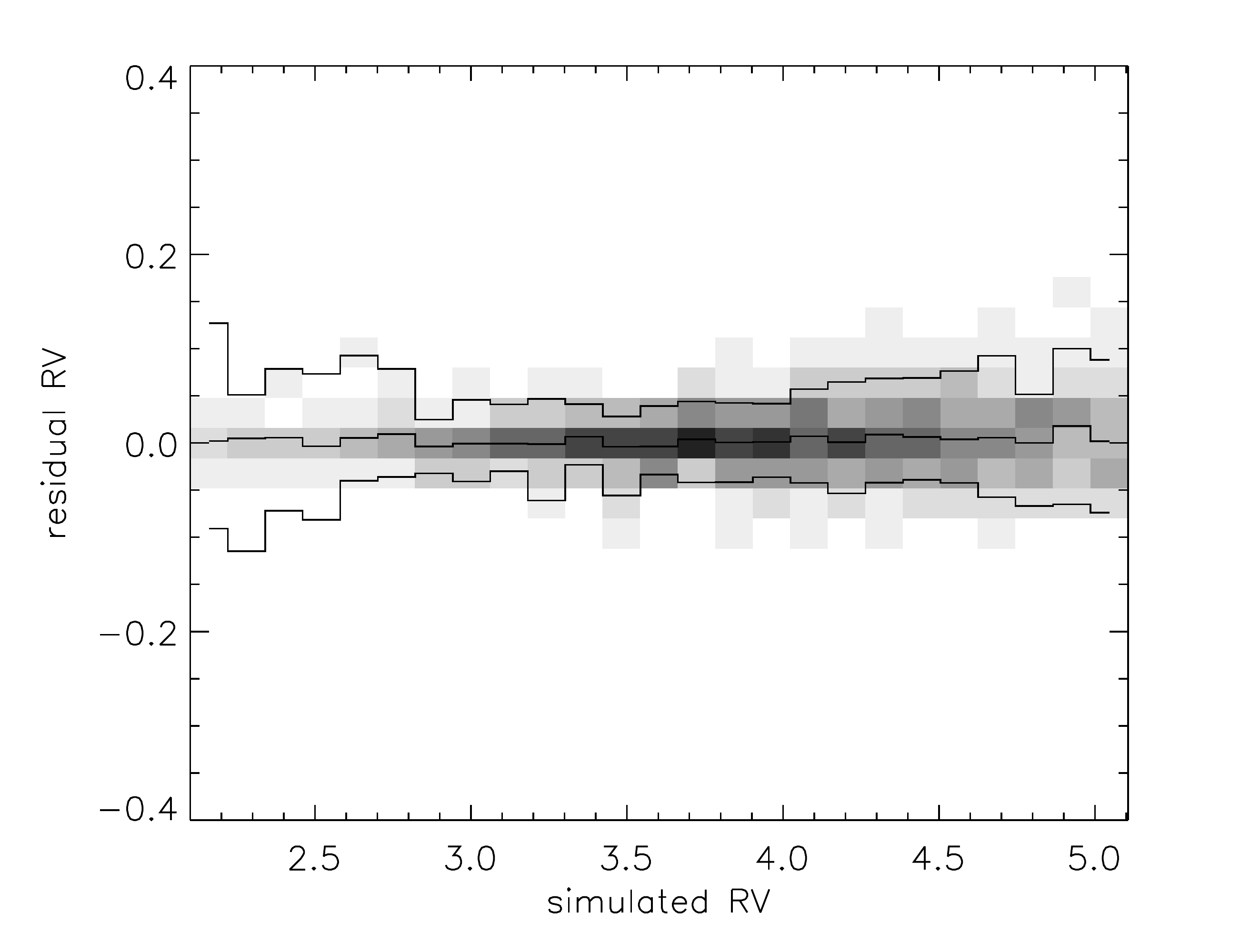}  
   \includegraphics[height=3.3in,angle=90]{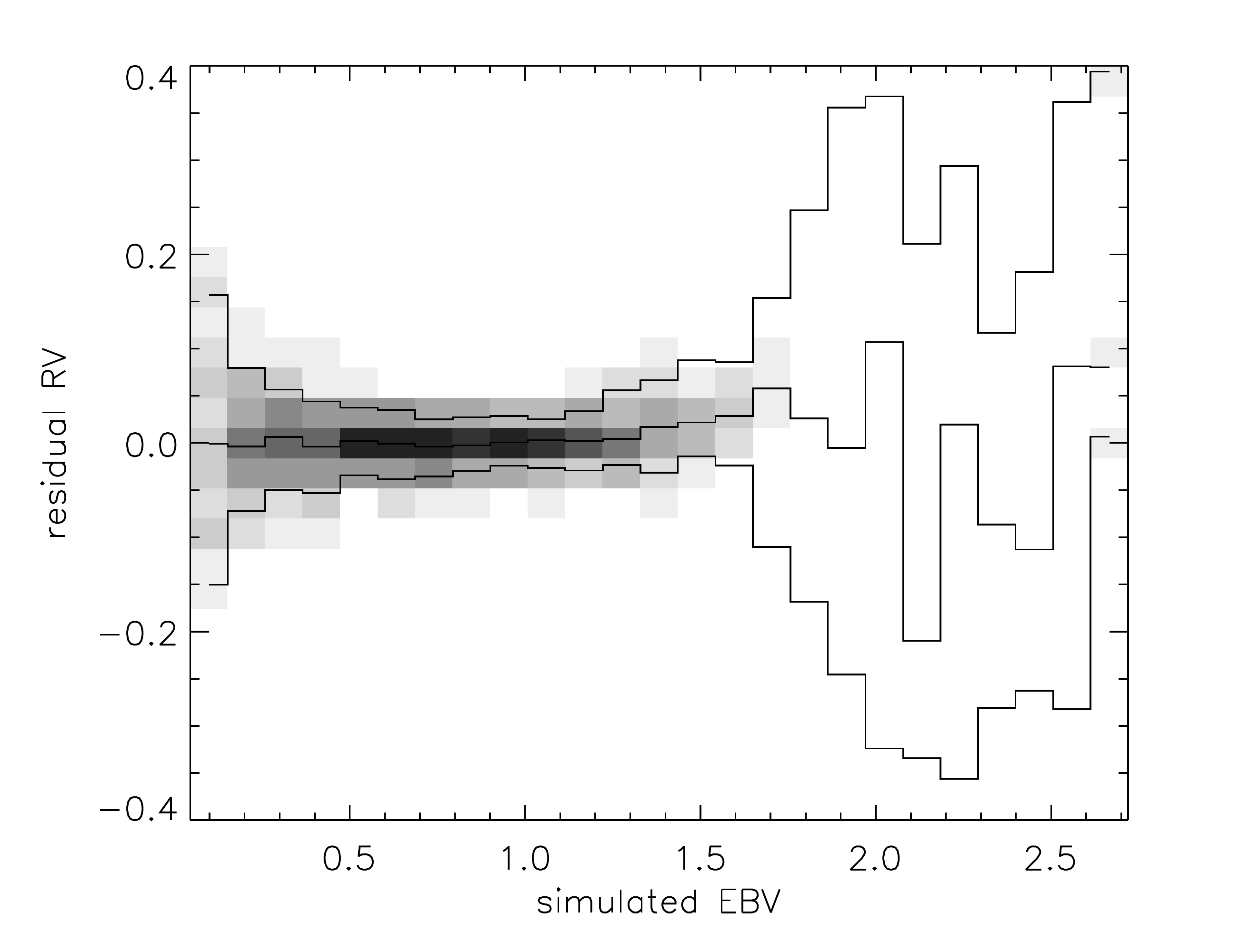}  
   \caption{$R_V$ residuals vs. $R_V$ (top panel) and vs $E(B-V)$ (bottom) for a single cloud along a line of sight. The grayscale squares show the density of pixels (single mock runs) falling in each 2D bin, relative to all other bins in the same column (i.e. bins with the same range of simulated parameter values). The squares being lighter in the leftmost columns of the top panel thus implies that samples with simulated $R_V < 2.5$ are spread out over a much wider range of residual $R_V$ values. The solid lines denote the $16$th, $50$th, and $84$th percentiles.
	This is the ideal case for the locus-shift method. We expect to have excellent precision in this limit. In general, there is a slight bias introduced by a selection effect where bluer stars and stars behind dust with lower $R_V$ are cut before other stars owing to the $g$ band dropping below threshold. This causes the estimated extinction to be grayer and results in a small positive residual. This effect is well below the typical uncertainty in our $R_V$ posteriors. Above an $E(B-V)$ of 1.5 we begin to lose a majority of the stars in a pixel. We recommend careful inspection of the uncertainties and quality factor before using such pixels.\\} 
\label{fig:mockred}
\end{figure}

\begin{figure}
   \includegraphics[height=3.3in,angle=90]{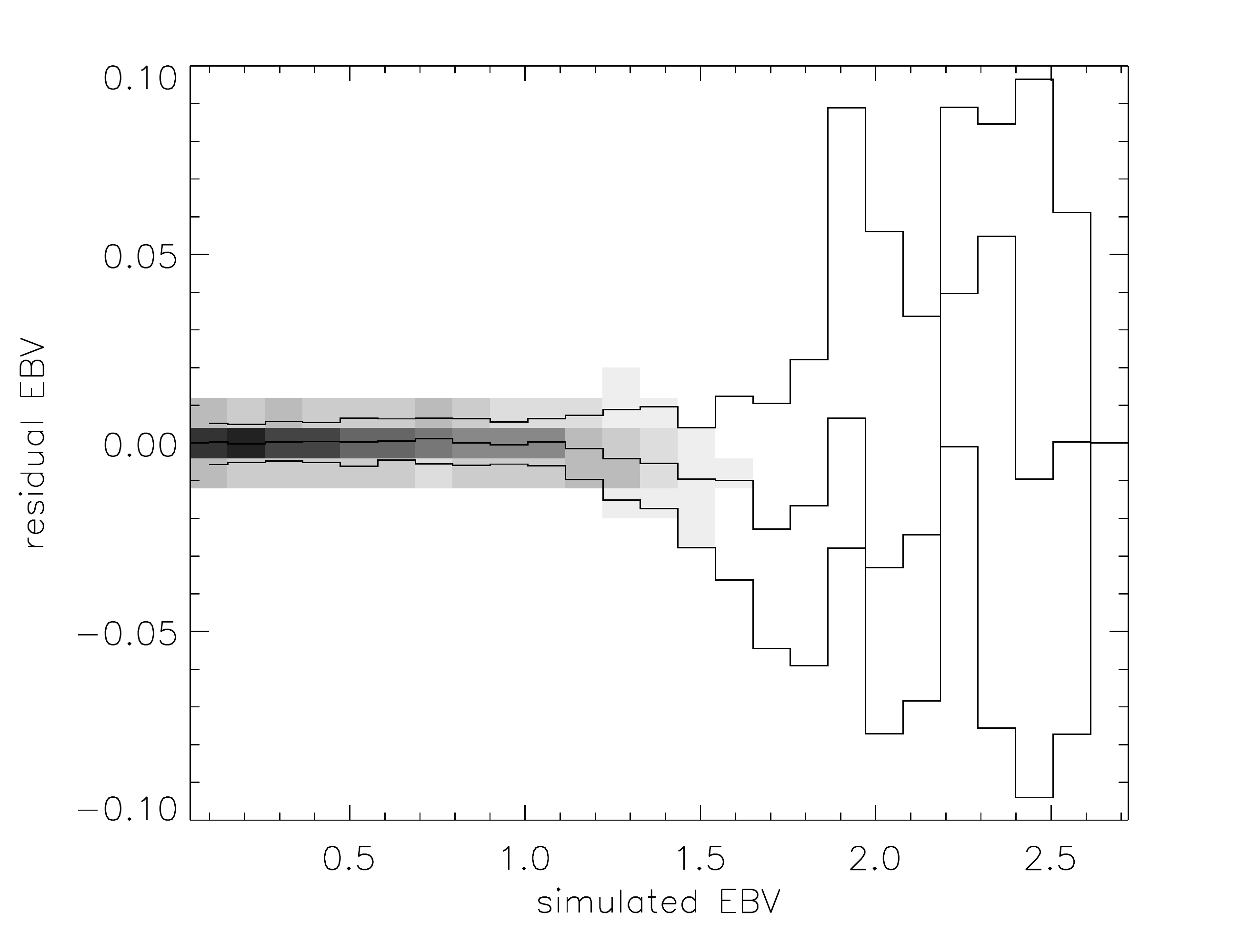}  
   \includegraphics[height=3.3in,angle=90]{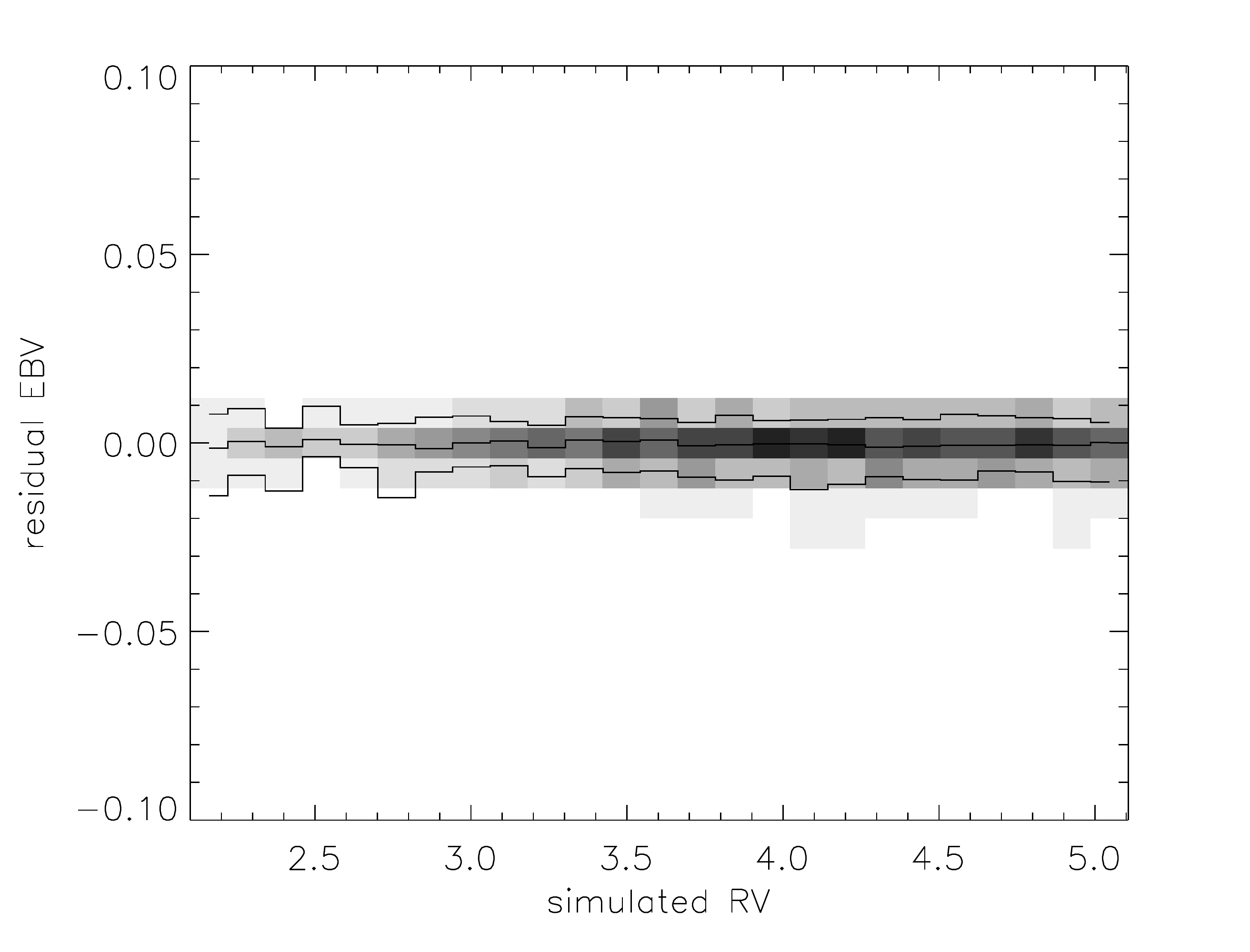}  
   \caption{$E(B-V)$ residuals vs. $E(B-V)$ and $R_V$. This is also for the ideal single-cloud case shown in Figure \ref{fig:mockred}. We see that our fits are expected to be quite precise (a spread of around $0.007$ mag on $E(B-V)$), but the accuracy suffers owing to selection bias at large values of $E(B-V)$. \\}
\label{fig:mockredE}
\end{figure}

\begin{figure}
   \includegraphics[height=3.3in,angle=90]{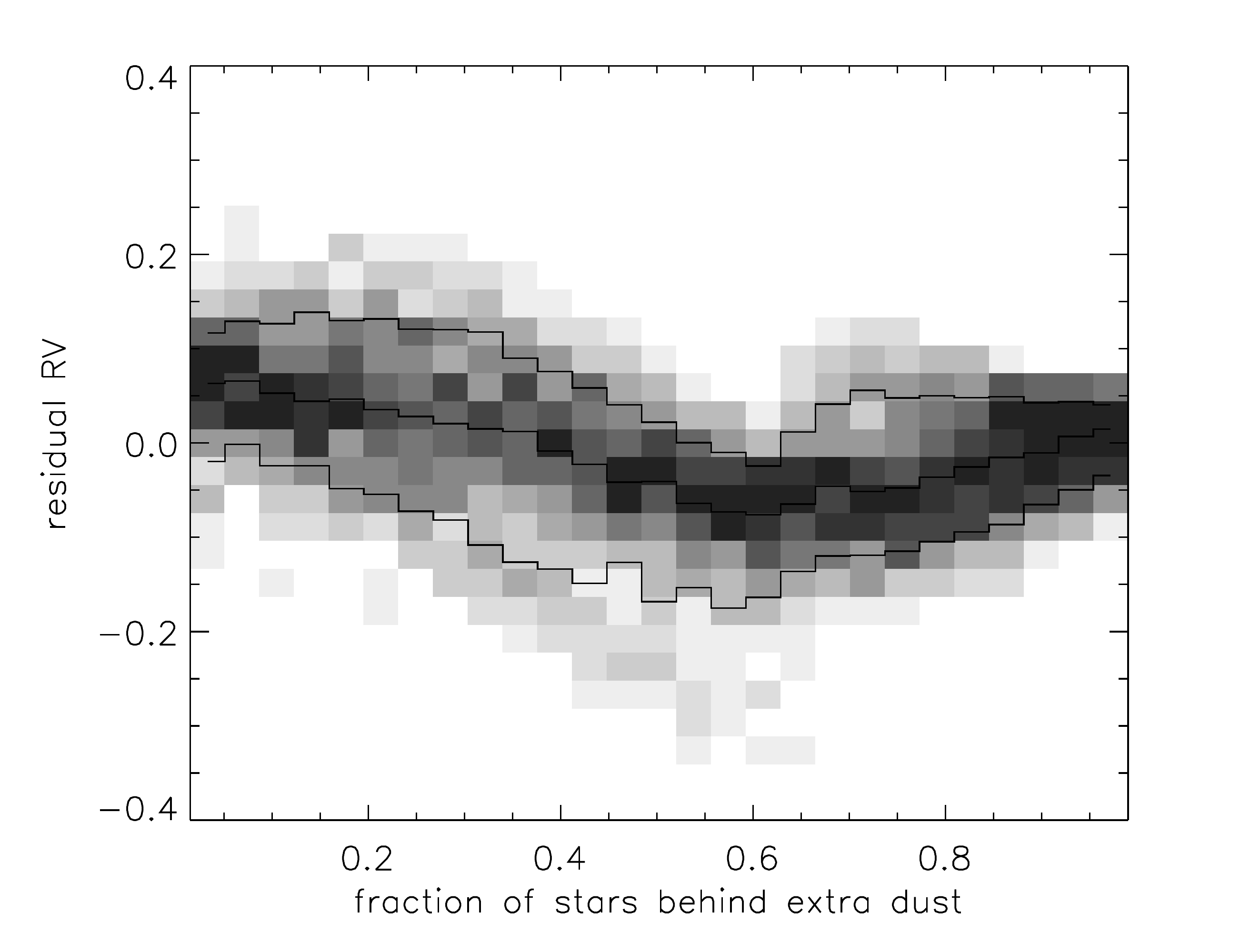}  
   \caption{$R_V$ residuals vs. the fraction of stars in a pixel behind a thicker cloud. These mock runs were set up to simulate what happens when a single pixel has a lot of subpixel spatial variation in dust column. As expected, we get the greatest residuals at a maximal mixing of $0.5$.\\}
\label{fig:mockfrac}
\end{figure}

\begin{figure}
   \includegraphics[height=3.3in,angle=90]{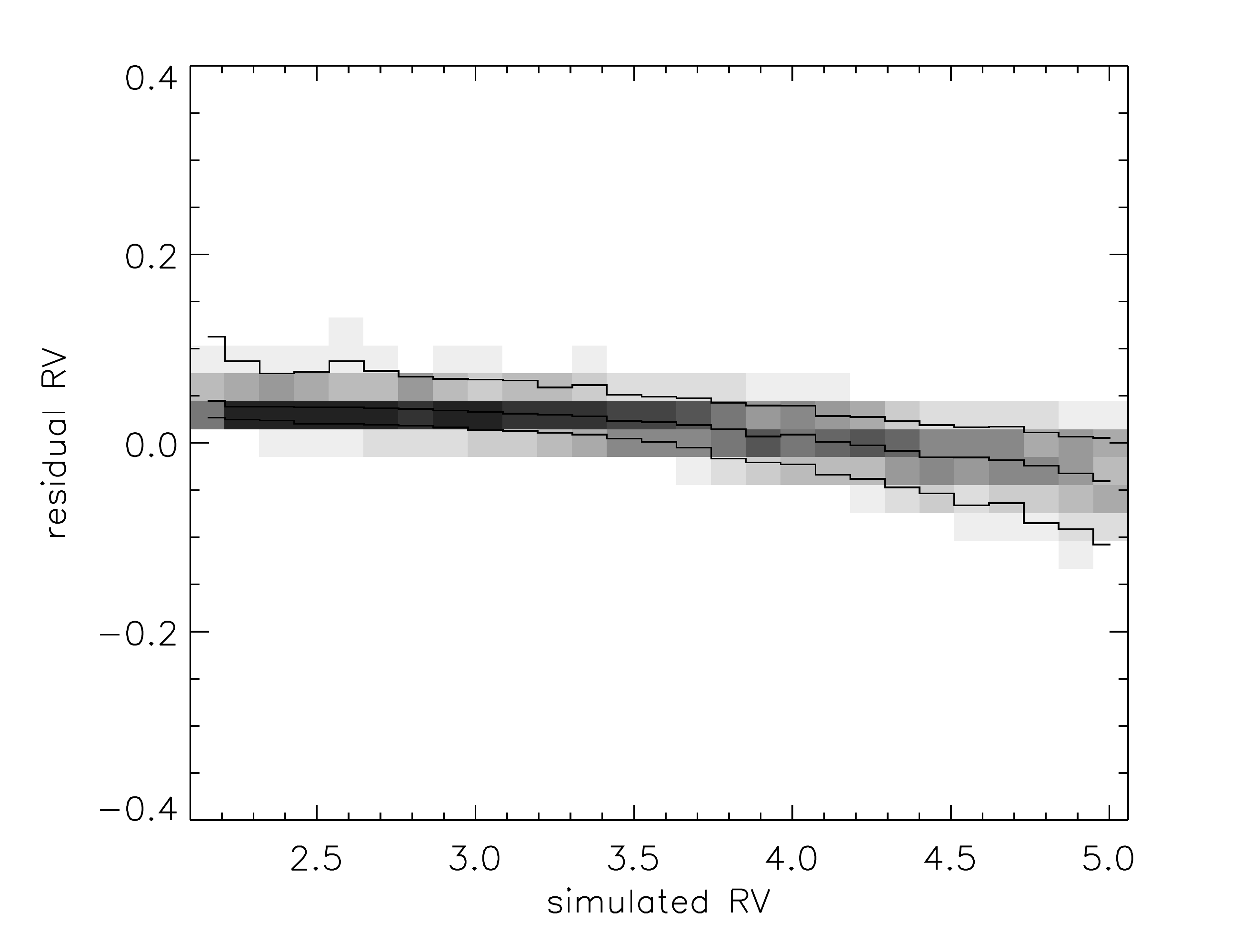}  
   \includegraphics[height=3.3in,angle=90]{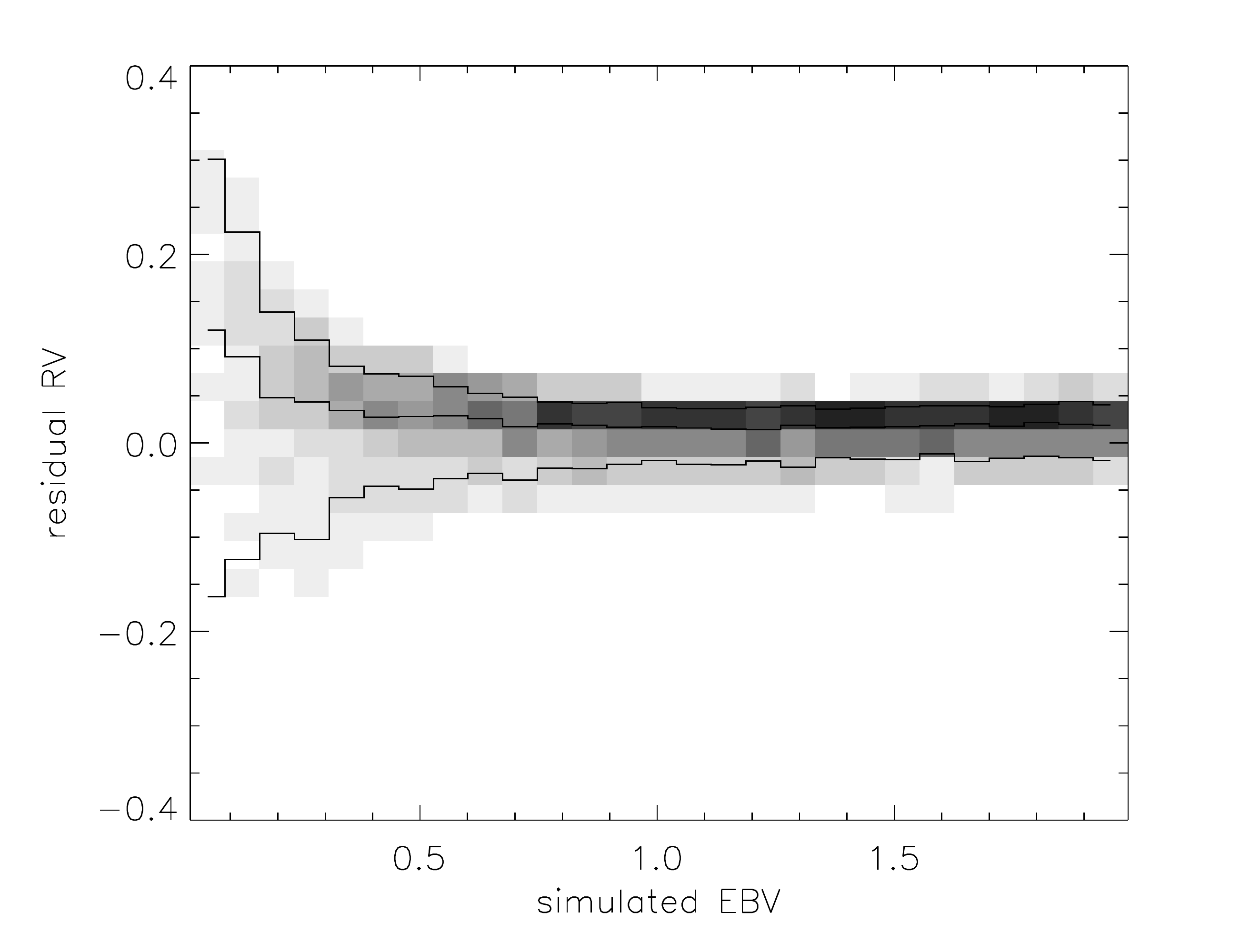}  
   \caption{$R_V$ residuals for pixels with two clouds along the line of sight. The horizontal axes are respectively the simulated $R_V$ and $E(B-V)$ for the closer cloud. These plots show that we expect our model to be reasonable even when stars are interspersed along multiple clouds along a line of sight. This is because clouds tend to come in discrete chunks (relative to interstellar distances) and stars behind additional clouds either are cut or are much fainter and thus have less influence on our fits. In this sense our maps of reddening are for the closest clouds along any given line of sight. This, however, does not hold for regions with continuous and nearly smoothly increasing reddening along a line of sight, such as near the Galactic plane. \\}
\label{fig:mockdouble}
\end{figure}

\begin{figure}
   \includegraphics[height=3.3in,angle=90]{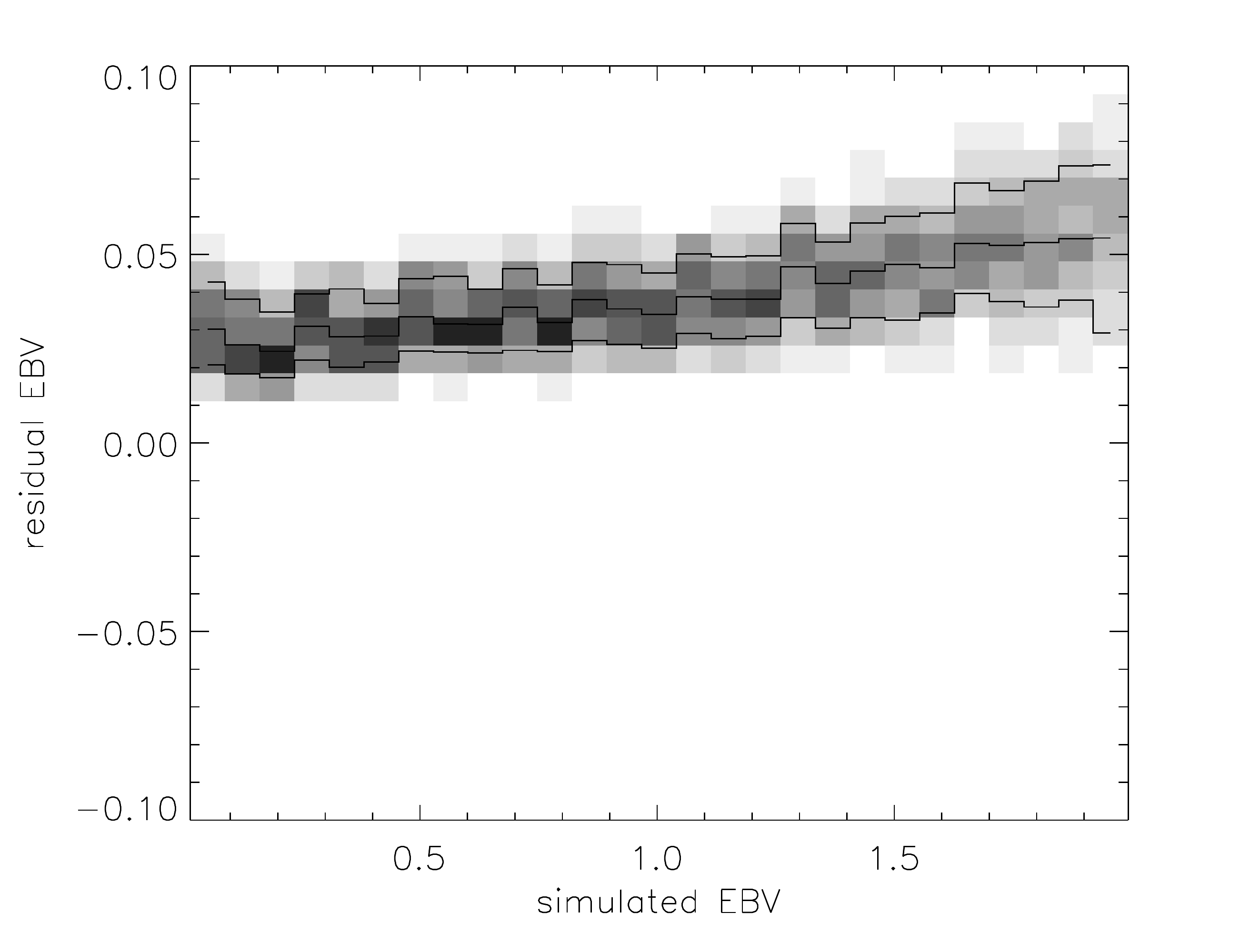}  
   \includegraphics[height=3.3in,angle=90]{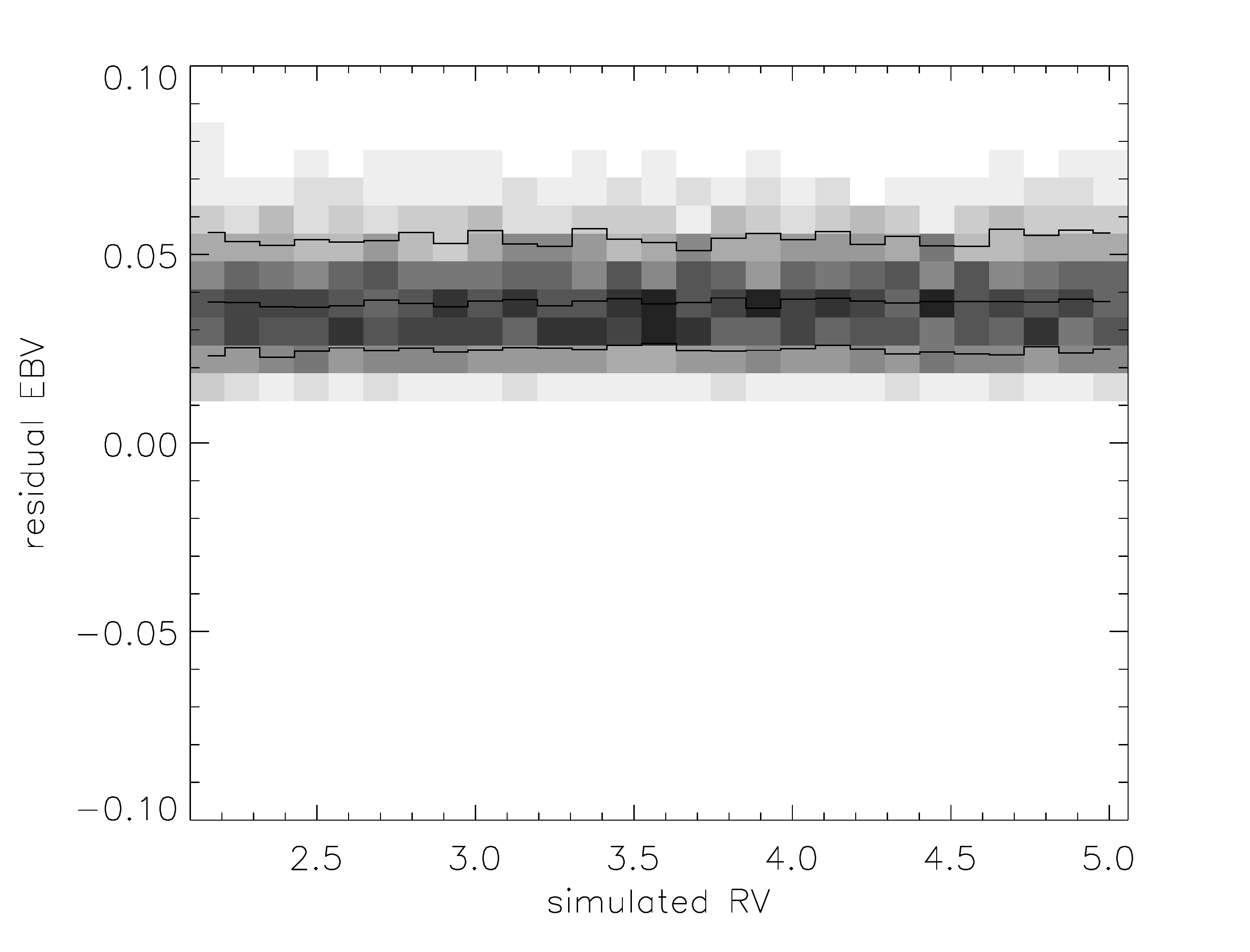}  
   \caption{$E(B-V)$ residuals vs. simulated $R_V$ and $E(B-V)$ for pixels with two clouds along the line of sight. Once again, we find residuals with the closer cloud to be small. \\}
\label{fig:mockdoubleEBV}
\end{figure}

\subsection{Mock Scenario: A Single Cloud}

Figures \ref{fig:mockred} and \ref{fig:mockredE} show how the residuals look for a single cloud - i.e. a discrete sudden increase in reddening - along a line of sight. This corresponds to the simple model used in the locus-shift method and thus should yield the smallest residuals. The stellar colors have been generated for a range of $R_V$ and $E(B-V)$ values for the mock dust cloud, and in the plots we show the marginalized distribution of residuals for each of the reddening parameters. There is a slight bias introduced by a selection effect where redder stars and stars behind dust with lower $R_V$ are cut before other stars owing to the $g$ band dropping below threshold. This causes the estimated extinction to be grayer and results in a positive residual.  Nevertheless, with these mock runs we are able to reproduce the 'true' $R_V$ to within $0.05$, which is much less than the typical uncertainty of $0.2$ in our $R_V$ posteriors. We determine $E(B-V)$ to within $0.007$, an order of magnitude less than the uncertainties in our $E(B-V)$ posteriors.

\subsection{Mock Scenario: Subpixel Variation}

In Figure \ref{fig:mockfrac} we show how our fits respond to a pixel with subpixel variation in dust column density. For this scenario we have two populations of stars, one behind dust with a column density corresponding to $E(B-V) = 0.2$ and another behind dust with $E(B-V) = 0.8$. We generate stellar colors for a range of $R_V$ values and also let the fraction of stars in the latter population $x_f$ vary from $0$ to $1$.  We plot the marginalized distribution of residuals as a function of $x_f$. The majority of residuals are less than $0.1$ $R_V$, once again less than the uncertainty in our $R_V$ posteriors. 
This is an intentionally pathological test case. Most pixels have smoother variation in extinction, even in the densest cores.

\subsection{Mock Scenario: Multiple Cloud Layers}

More typical pixels have low subpixel spatial variation in extinction and instead have stars interspersed among dust clouds at multiple distances along a line of sight. This means that some stars in a pixel have additional reddening owing to being behind more clouds.  We generate stellar colors for this scenario by first reddening half the stars in a pixel with a fixed reddening vector of $R_V = 3.1$ and $E(B-V) = 0.2$ and then reddening all the stars (including the half already reddened) with an additional reddening vector of varying $R_V$ and $E(B-V)$. This models the effect of one nearby cloud of varying reddening being in front of all stars in a pixel and an additional cloud of fixed reddening appearing at the median distance of the stars. In these situations, residuals in $R_V$ are typically around $0.05$ when we compare the estimated $R_V$ of the entire pixel to the simulated values for the closer cloud. Figure \ref{fig:mockdouble} shows the marginalized distribution of residuals as a function of the true $R_V$ of the closer cloud. We obtain the 'correct' answer for the closer cloud owing to selection bias. Stars behind extra dust are more likely to be cut or not observed owing to increased extinction. Additionally, those that do pass the cut are fainter than stars behind only the first cloud, and we expect the former to have less influence on our fits. Figure \ref{fig:mockdoubleEBV} makes the same comparison as a function of the true $E(B-V)$ of the closer cloud.

Altogether, we see that our fits should be accurate to $0.05$ in $R_V$ for most lines of sight where the reddenings are well approximated by a single dust cloud. They are also expected to be accurate to $0.1$ in $R_V$ even in some problematic lines of sight.

\begin{figure*}
\centering
	\includegraphics[width=\textwidth]{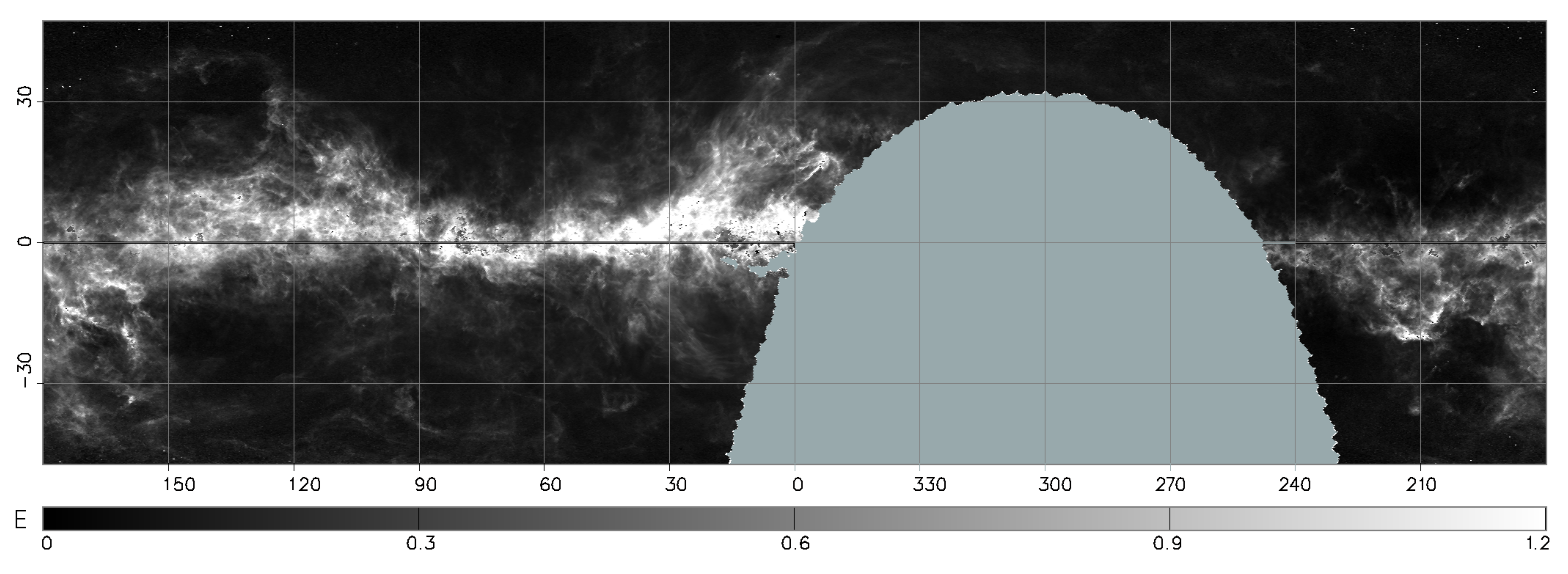} \\ 
  \includegraphics[width=\textwidth]{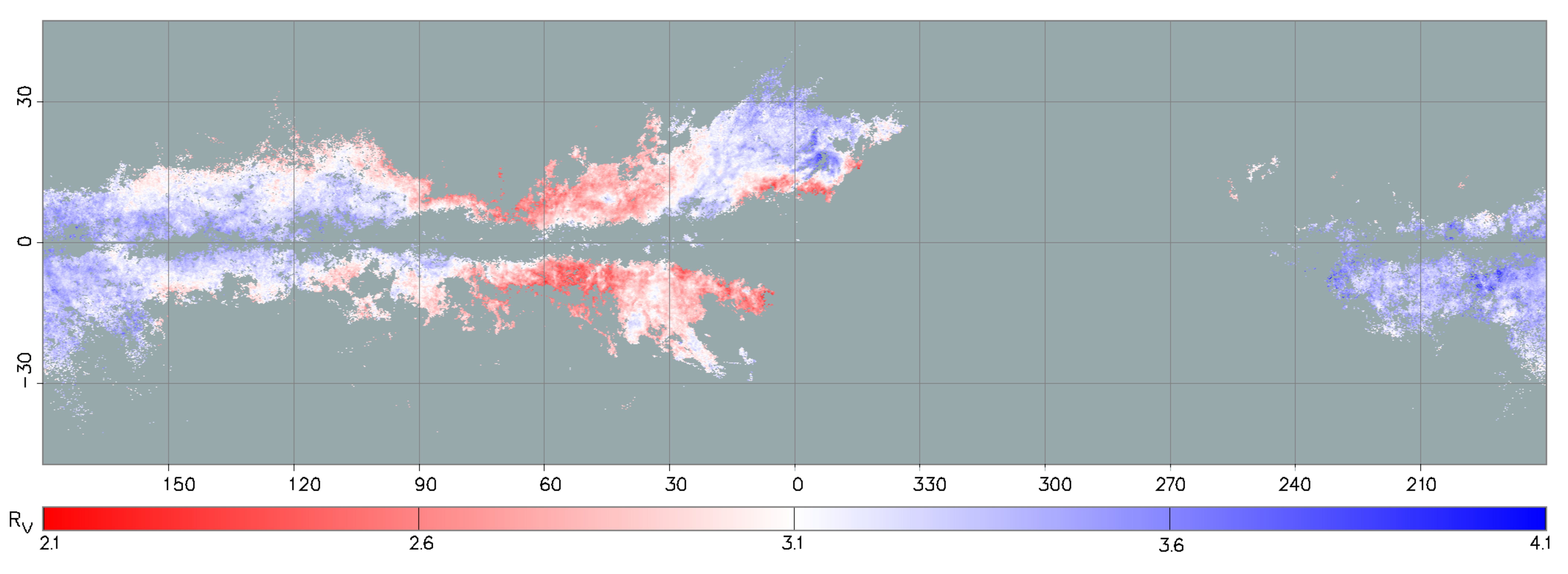} 
  \caption{Maps of $R_V$ and $E(B-V)$ using the locus-shift method. The top panel shows a map of $E(B-V)$ for values ranging from $0$ to $1.2$. Pixels with no data, or unphysical fits (such as negative $E(B-V)$), are denoted by a cool (bluish) gray.  The bottom panel is color-coded so that red pixels have $R_V < 3.1$ and blue pixels have $R_V > 3.1$. We pick $R_V = 3.1$ as neutral white owing to several past results claiming the mean $R_V$ in the Milky Way to be $3.1$ \citep{fitz}. As evident from this map, we estimate a mean $R_V$ slightly larger than this value. For the $R_V$ map we employ a hard cut on pixels with $E(B-V) < 0.2$ or quality factors less than $-2.2$, which roughly corresponds to having a total estimated $R_V$ uncertainty greater than $0.1$, including model uncertainties and systematics. These pixels are once again denoted by the cool gray. See Figure \ref{fig:RVvE_dist} for a binned scatterplot of these two maps. \\}
\label{fig:f7} 
\end{figure*}

\section{The Locus-shift $R_V$ Map}\label{sec:rvmap}

In Figure \ref{fig:f7} we present maps of $R_V$ and $E(B-V)$, converted from locus shifts as specified in Section \ref{sec:choose}. We also zoom in on interesting clouds in Figures \ref{fig:RVoph}--\ref{fig:RVori}. The maps are at $N_{side} = 256$ and run from $b=-45$ to $b=45$. In the $E(B-V)$ map, white corresponds to greater extinction, and the cool gray denotes pixels with no information. For the $R_V$ map, white corresponds to $R_V=3.1$, red is for $R_V$ values less than $3.1$, and blue is for values greater than $3.1$. We select pixels that are above a threshold in both $E(B-V)$ and a quality factor $Q$ related to the reliability of a pixel (as defined below). The $E(B-V) > 0.2$ threshold ensures that we are insensitive to unmodeled variations in the stellar population, and the $Q$ threshold ensures that the fits themselves are informative. On the other hand,	 for the zoomed-in maps we employ a value mask so that darker values correspond to less reliable pixels, with black denoting no available information. We choose the hard cut for the large map to make it easier to see the smoothly varying spatial trends in $R_V$ through the Galaxy. We choose the value mask for the zoomed-in maps to convey more information about pixel uncertainties. We also hope that these two different graphical representations demonstrate to the reader that the maps are only intended to give rough ideas of $R_V$ variations, and that for detailed comparisons it is best to use the full reddening vectors provided in our data release, along the related uncertainties.

\subsection{Accounting for Galactic Variation with the $E(B-V)$ Threshold}\label{sec:galvar}

Since stellar populations vary across the Galaxy, we must have a method for estimating or accounting for their effects on our fits.
Following the process outlined in Section \ref{sec:locus}, we construct a model locus for low galatic latitudes and another for high latitudes.  
This dual model assumes that stellar populations can be modeled by a thick disk and a halo component, and that all lines of sight of interest are some combination of these two. Since our targets avoid regions with high extinction (see Section \ref{sec:rvmap}), we do not need to model the thin disk or the bulge. We run our fits with both models and find that the two sets of results begin to agree with each other within uncertainties when the total extinction $E(B-V) > 0.2$. We claim that the thick disk and halo models represent the maximum possible variation in the stellar locus, and thus use $E(B-V) > 0.2$ to select pixels we do not expected to be dominated by model uncertainties. 

We note that the $E(B-V)$ threshold of $0.2$ is greater than the value of $0.06$ required to be insensitive to variations in the unreddened locus because differences in colors for different stellar populations are greater than those introduced by the de-reddening procedure. We therefore use the larger value for all analyses below, unless otherwise stated. Furthermore, given that we are using the proxy definition ${R_V}_{gy}$ (from Section \ref{sec:rvdef}), the threshold allows us to avoid biases that may arise from having a small value of $E(g-r)$ in the denominator.

\subsubsection{Accounting for Fitting Uncertainties with the Quality Factor}\label{sec:qfactor}

The reliability of a pixel is based on a quality factor $Q$ that depends on the evidence per star and the uncertainty in $R_V$, defined as follows:
\begin{equation}
Q = \log{(Z)}/N - A\sigma_{R_V},   \\
\label{eq:qfactor}
\end{equation}
where $Z$ is the evidence, $N$ is the number of stars, and the constant $A$ is chosen so that $\log{(Z)}/N$ and $-A\sigma_{R_V}$ have roughly the same standard deviation as each other for a pixel well-described by our locus-shift model. We find $A=0.69$ to be a reasonable choice, but note that this number is flexible. For example, if we were to use a different parameterization for $R_V$, a different value of $A$ may be optimal.  Although we recommend a cut at $Q=-2.2$ for rough comparisons, the best method for estimating confidence in a fit would be to determine whether the evidence or $R_V$ standard deviation (or some other statistic of the $R_V$ posterior) is most relevant for the particular use case. The formula above has been optimized primarily for visual inspection of reddening maps. For plotting purposes, we convert $Q$ to
\begin{equation}
M = 0.5 + 0.5\tanh{ \left[ (Q-\text{median}(Q))/\sigma_Q \right] },  \\
\label{eq:mfactor}
\end{equation}
which has a more even distribution from $0$ to $1$ and is intuitively similar to the evidence multiplied by a probability related to the uncertainty in $R_V$. This makes $M$ well suited for masking out Healpix pixels based on our confidence in the fit parameters.

As expected, $Q$ masks out regions with few stars or low reddening, since there is not enough information for a good fit. It also masks out regions with a lot of continuous reddening along the line of sight (instead of discrete clouds) because the stars in the pixel have been attenuated by too wide a range of reddening vectors, and the locus-shift model is a bad approximation of the data. Such cases occur predominantly around the Galactic plane.

\subsection{Accessing the Map}

The locus-shift map can be downloaded as a FITS file from the Harvard Dataverse via the link \url{http://dx.doi.org/10.7910/DVN/TJGJWW}. The structure of the data is as described in Table \ref{tab:fits}:

\begin{table}
	\caption{Summary of Keys in Reddening Map FITS File}
	\label{tab:fits}
  \begin{tabular}{ | l | r | r | }
    \hline
    Key (Tag) & Format & Description  \\
		\hline
    healpix            & $2\times$\text{int64}     & (Healpix, $N_{side}$) \\ 
    nstars             & \text{int64}              & no. of stars \\ 
    locus\_shift\_mean & $4\times$\text{float64}   & mean locus-shift (LS) \\
    locus\_shift\_best & $4\times$\text{float64}   & best LS \\
		covariance         & $4^2\times$\text{float64} & covariances of LS \\
    logZ               & \text{float64}            & log of evidence \\ 
    Gelman\_Rubin      & $4\times$\text{float64}   & GR diagnostic of LS \\ 
    RV\_EBV\_F99       & $4\times$\text{float64}   & ($R_V$, $\sigma_{R_V}$, $E$, $\sigma_{E}$) \\
    RV\_EBV\_PC        & $4\times$\text{float64}   & ($R_V$, $\sigma_{R_V}$, $E$, $\sigma_{E}$) \\
    RV\_proxy          & $2\times$\text{float64}   & ($R_V$, $\sigma_{R_V}$) \\
		Qfactor            & $3\times$\text{float64}   & ($Q$, $M$, $M'$) \\
    \hline
  \end{tabular}
\end{table}

\begin{figure}
   \leavevmode\epsfxsize=7cm\epsfbox{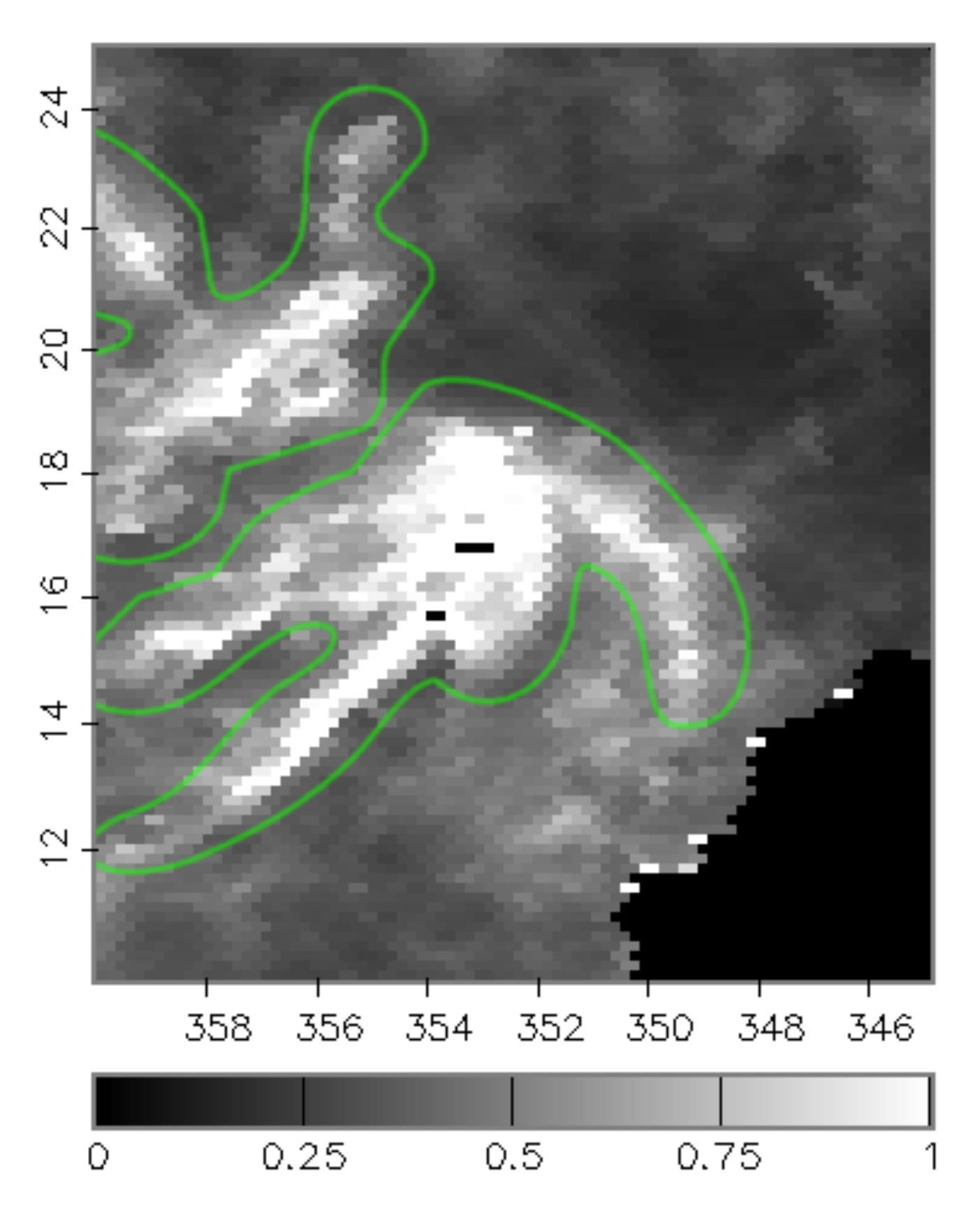} 
   \leavevmode\epsfxsize=7cm\epsfbox{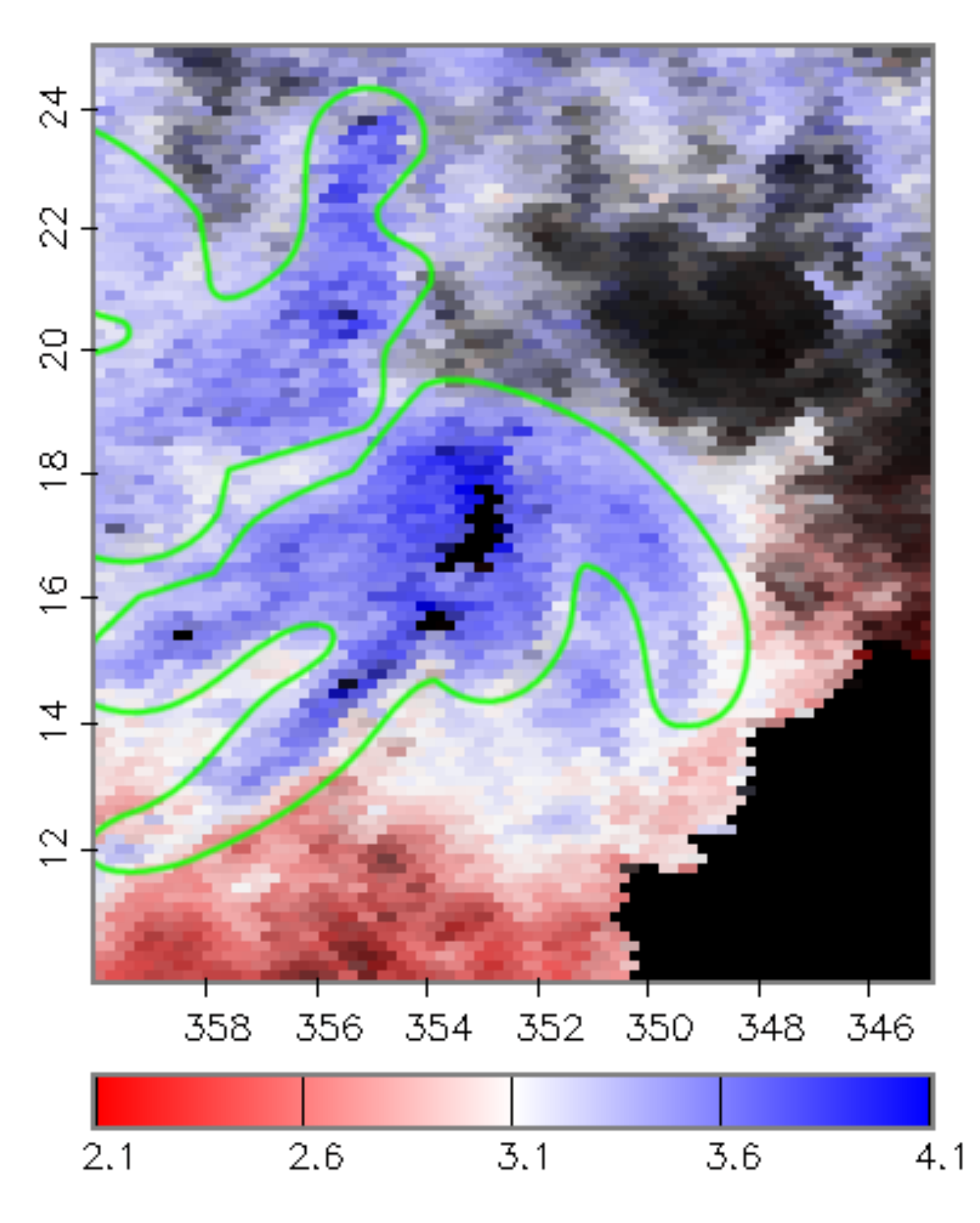} 
   \caption{Top panel: map of the $E(B-V)$ variation in Ophiucus. Bottom panel: variation in the expected value of $R_V$ for the same region, with green contours around clouds to aid comparison. For our zoomed-in maps of individual clouds we elect to color-code the information a little differently than in Figure \ref{fig:f7}. The degree of saturation in red and blue still corresponds to low and high $R_V$ values. However, we now also let the value, i.e. the lightness, of the pixel increase monotonically with the adjusted masking fraction $M'$ (Eq. \ref{eq:MME}). Completely black pixels have zero information. \\
	As evidenced by comparing the two panels, $R_V$ is highly correlated with $E(B-V)$ in this cloud, which corroborates some theories about grain size distributions in dense clouds. On the other hand, the correlation may just imply that our parameters have artificial dependencies. Figure \ref{fig:RVcep} proves that this is not the case, and that we are independently sensitive to both $R_V$ and $E(B-V)$. \\ \\}
\label{fig:RVoph} 
\end{figure}

\begin{figure}
   \leavevmode\epsfxsize=8cm\epsfbox{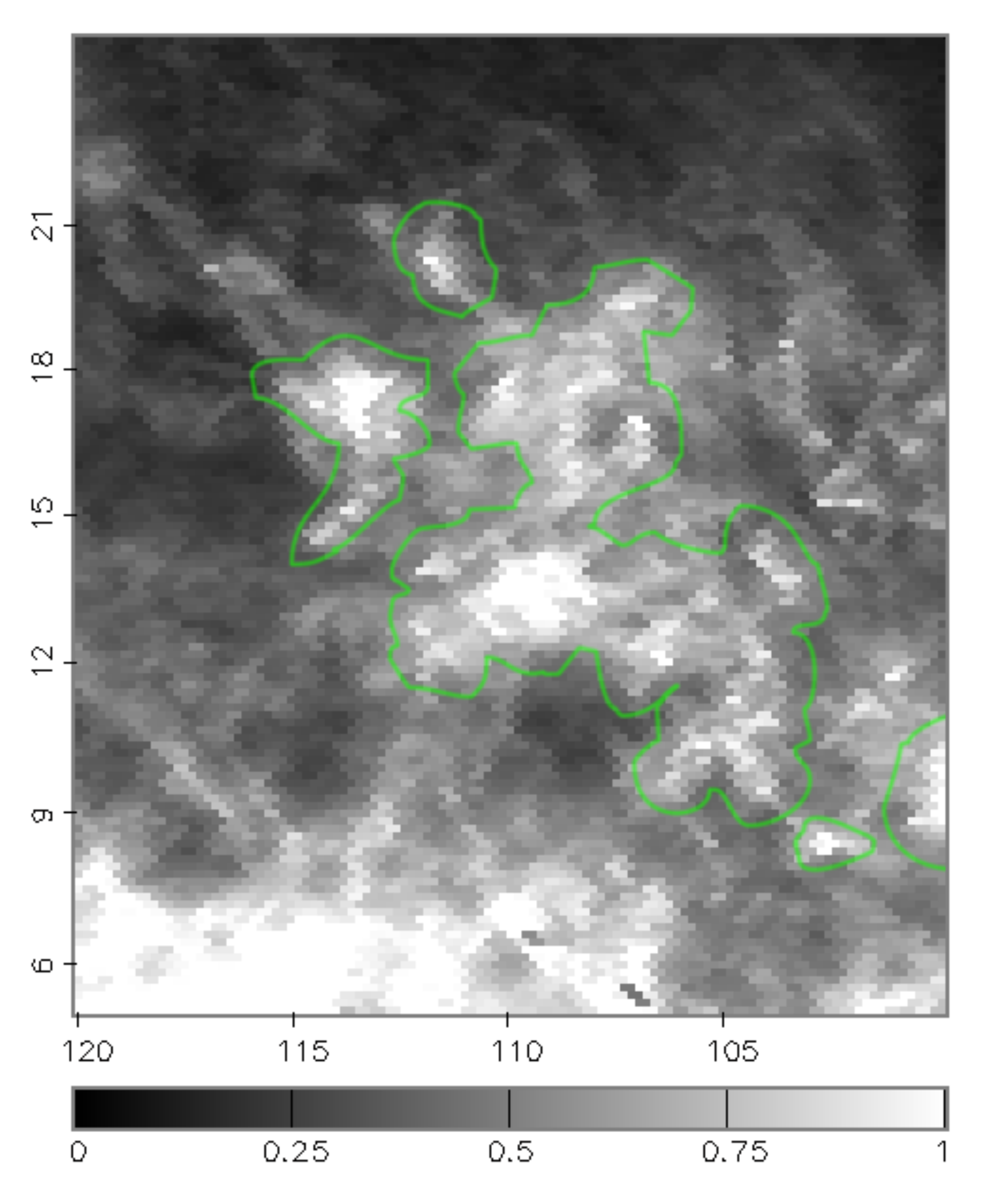} 
   \leavevmode\epsfxsize=8cm\epsfbox{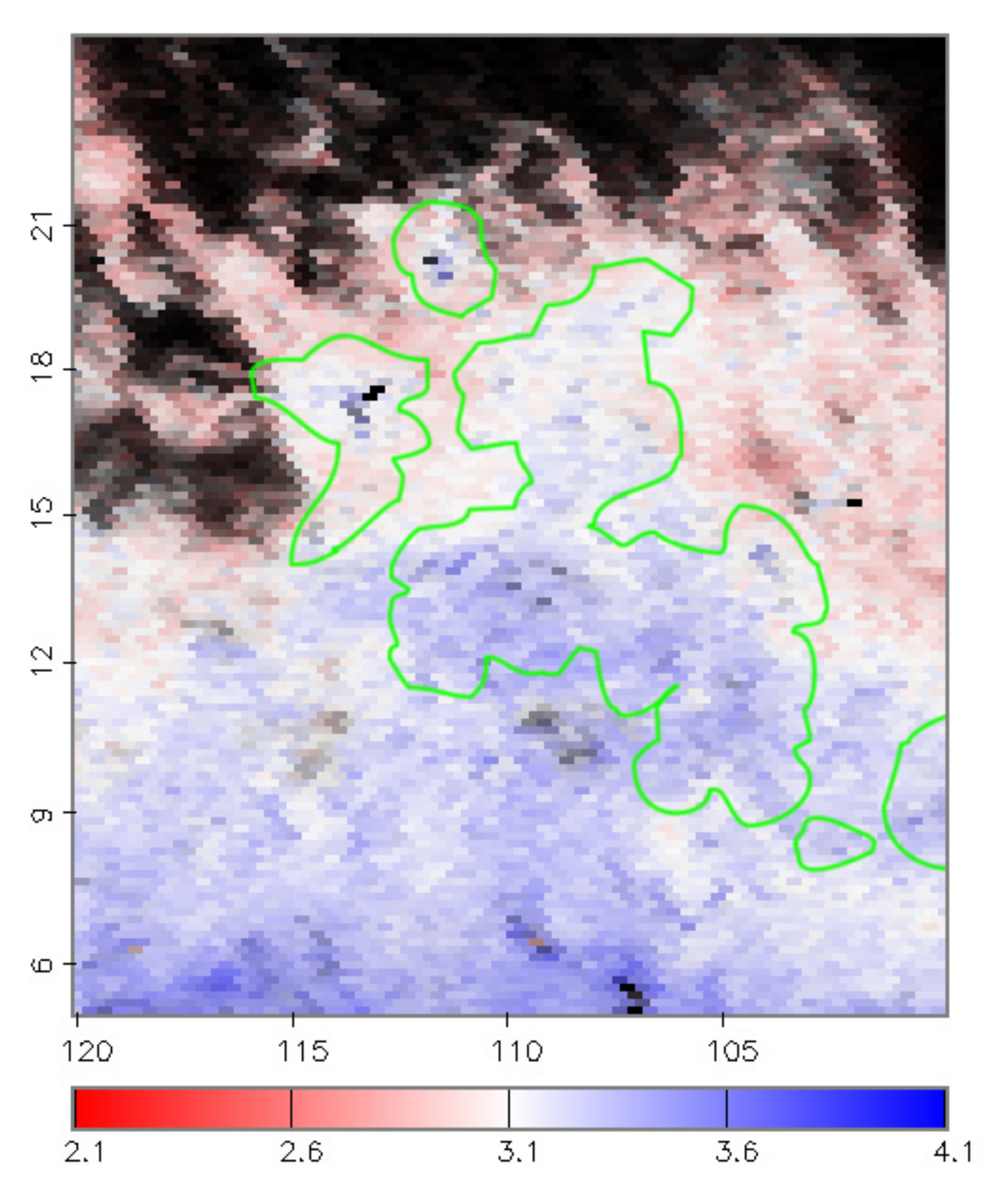} 
   \caption{Top panel: map of the $E(B-V)$ variation in Cepheus. Bottom panel: map of the $R_V$ variation in Cepheus. Here we see very little correlation between $R_V$ and $E(B-V)$. There is minor correlation in the dense regions of the cloud around $b=18	$, which still implies that larger grains do form there. However, the other pixels show $R_V$ varying independently of $E(B-V)$, which suggests that we have the potential to probe other properties of dust, e.g. perhaps its chemical composition. Of course, other factors, such as a smoothly increasing dust column, may influence our determination of $R_V$. \\
Additionally, the lack of correlation in Cepheus may have been due to changing stellar populations as we move closer to the galactic disk, but the next two Figures \ref{fig:RVper} and \ref{fig:RVori} show that $R_V$ exhibits behavior contrary to this as well. }
\label{fig:RVcep}
\end{figure}

\begin{figure}
   \leavevmode\epsfxsize=8cm\epsfbox{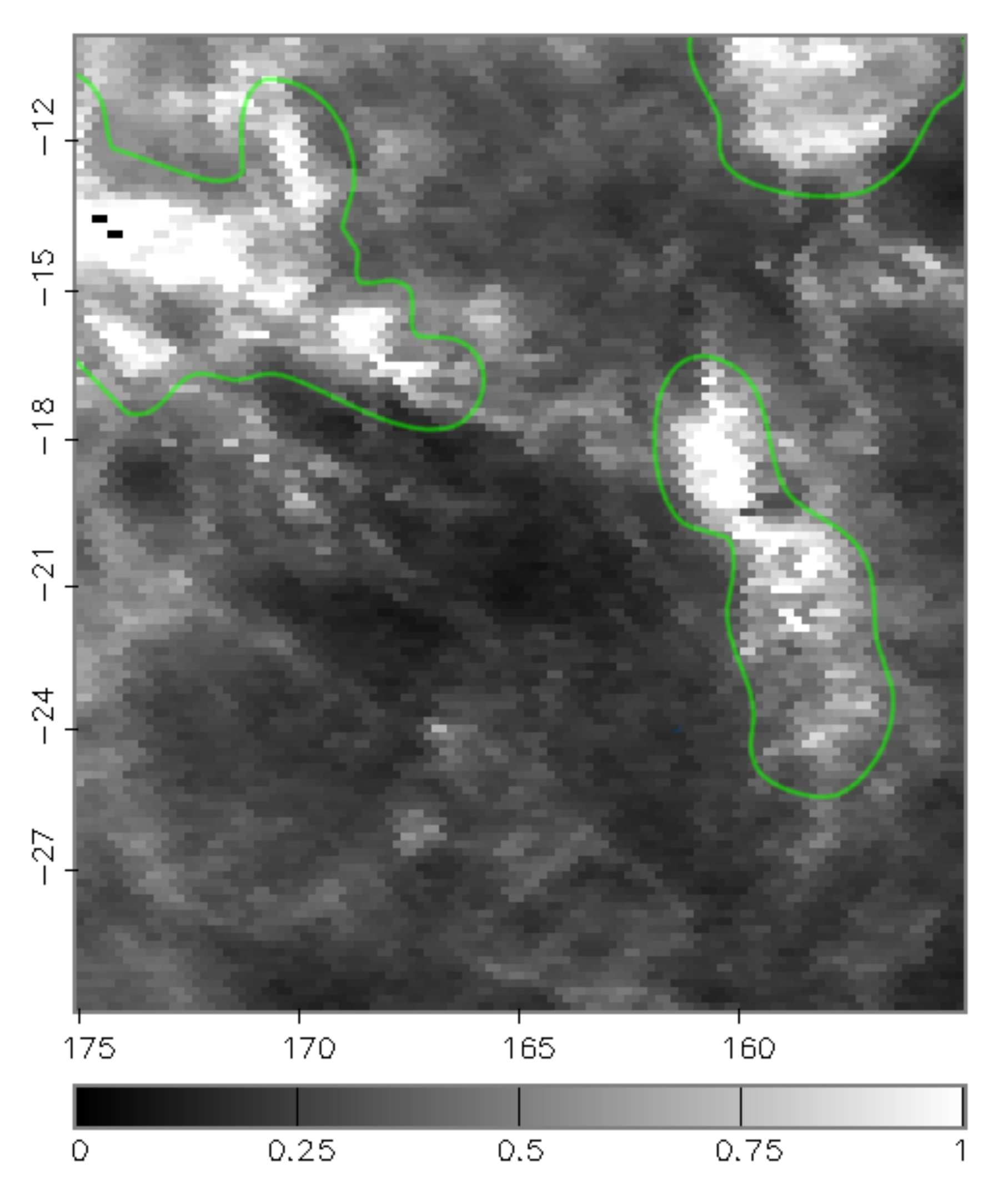} 
   \leavevmode\epsfxsize=8cm\epsfbox{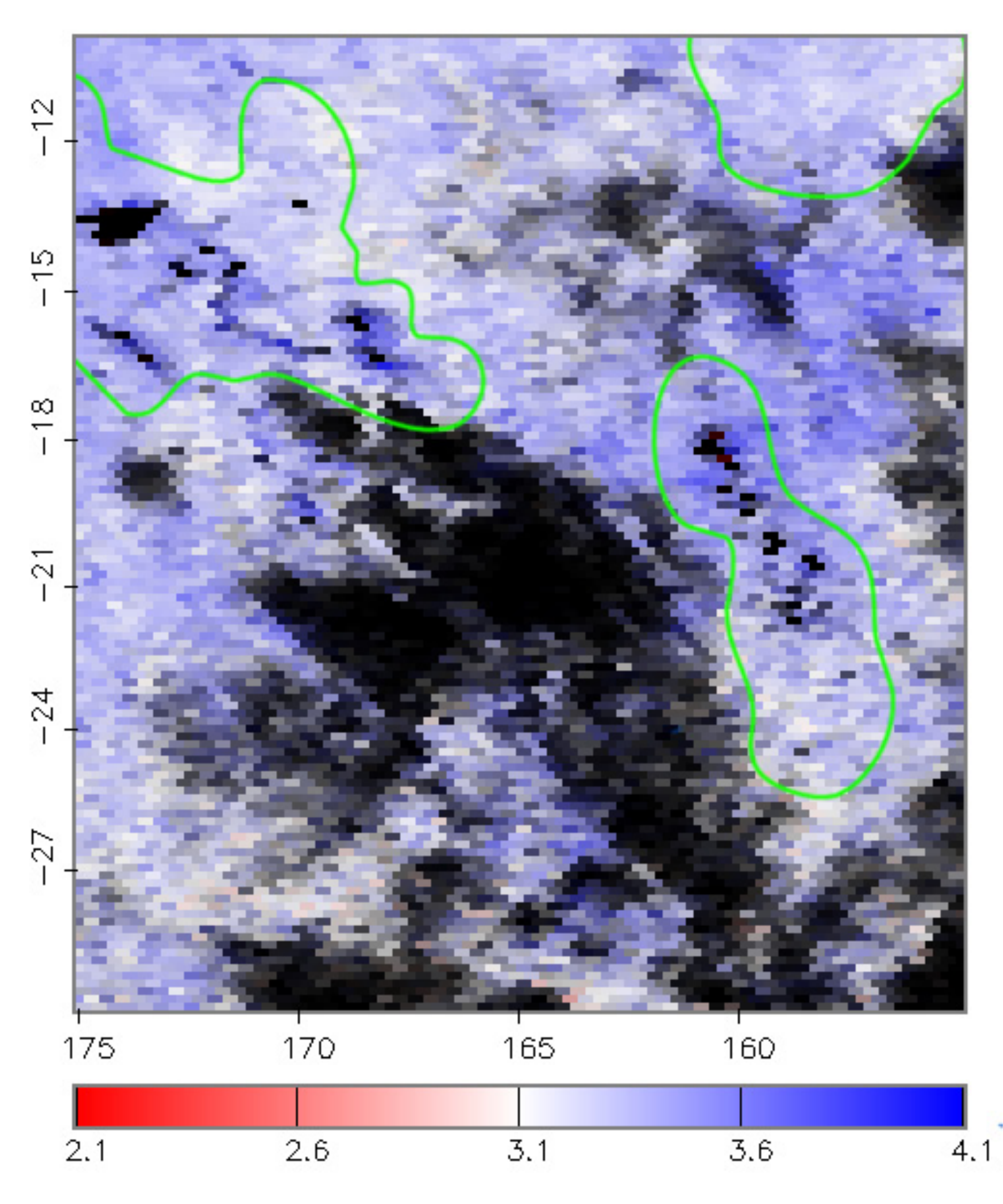}
   \caption{Maps of the $E(B-V)$ and $R_V$ variation in Perseus and Taurus. We show contours around dense clouds to facilitate comparisons between features in the top and bottom panels. Once again, we see a correlation between dense cores and high $R_V$. However, the relation is not as pronounced as in Ophiuchus, and additionally there are patches with the opposite relation - i.e., low column densities correlated with higher $R_V$ (e.g. around $l=-15$, $b=160$). This gives us confidence that we are probing a wide range of dust properties. \\	}
\label{fig:RVper}
\end{figure}

\begin{figure}
   \leavevmode\epsfxsize=8cm\epsfbox{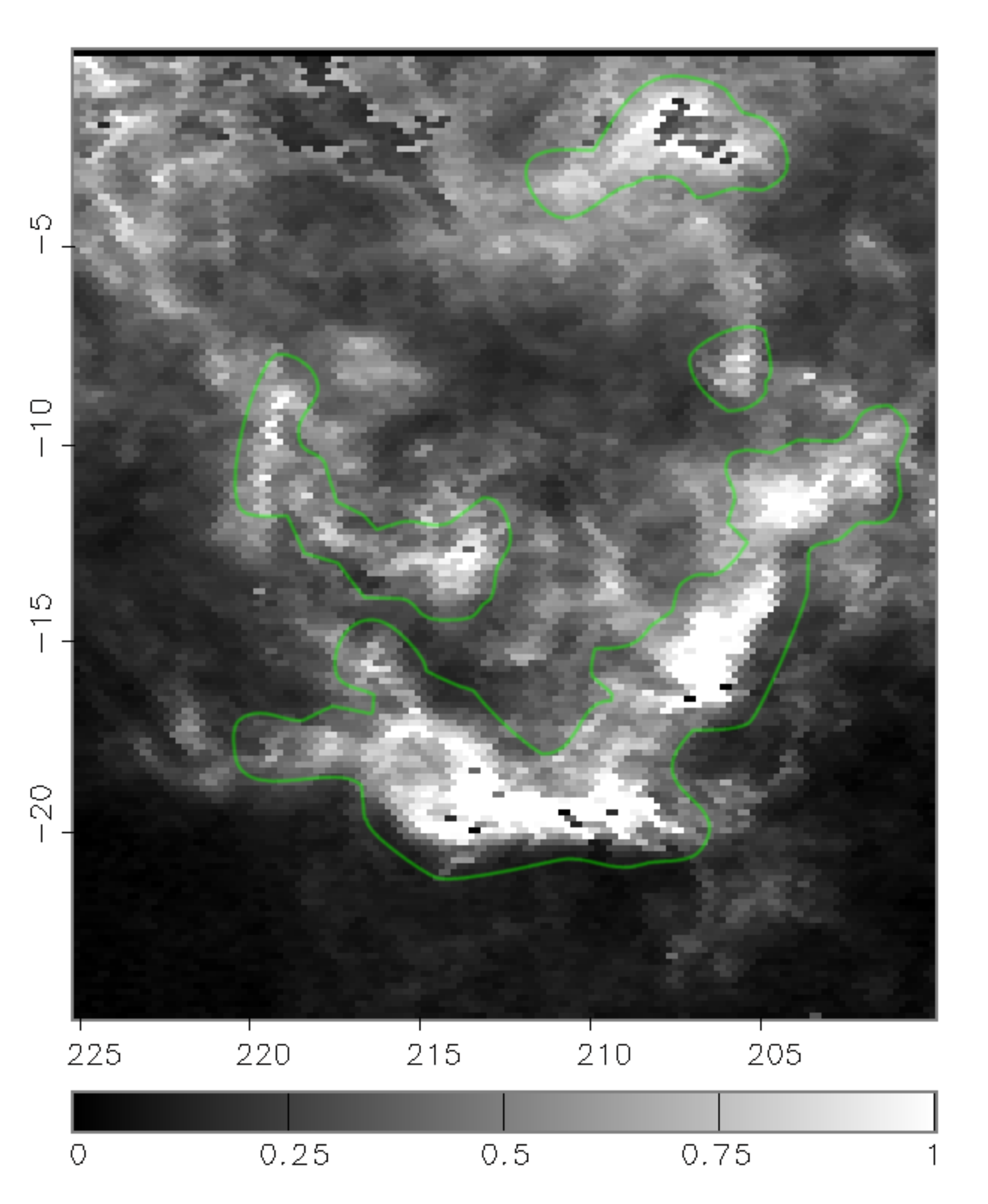} 
   \leavevmode\epsfxsize=8cm\epsfbox{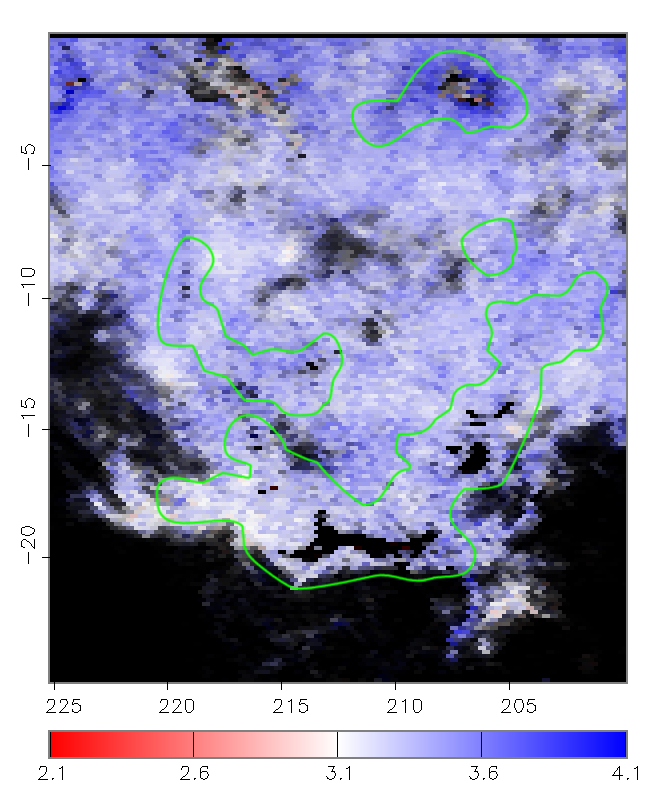} 
   \caption{Maps of $E(B-V)$ variation (top panel) and $R_V$ variation (bottom panel) in 	n. We see that, similarly to Cepheus and Perseus, Orion does not show a strong correlation between $E(B-V)$ and $R_V$. In fact, some pixels in dense clouds are not as blue as surrounding pixels, indicating a small but negative correlation between $E(B-V)$ and $R_V$. The wide range of correlations found in the distinct clouds in these figures suggests that we are measuring real variations in reddening.\\	}
\label{fig:RVori}
\end{figure}

In the table, $E$ is shorthand for $E(B-V)$.  Every element or row in the first data unit has the ten keys listed above. Each element corresponds to a Healpix pixel for which we have data. The content of each key is summarized above.  \texttt{locus\_shift\_mean} and \texttt{locus\_shift\_best} are $4$-element double precision arrays of the mean locus-shift and the best locus-shift sample, respectively. The $4$ elements correspond to the $g-r$, $r-i$, $i-z$, and $z-y$ color-shifts. \texttt{covariance} is the full $4\times4$ covariance matrix of color-shifts. \texttt{RV\_EBV\_F99} has four elements corresponding to the mean $R_V$, mean $E(B-V)$, standard deviation of $R_V$, and standard deviation of $E(B-V)$ of the locus-shifts according to the F99 reddening law.  \texttt{RV\_EBV\_PC} has the same structure, except according to the principal component formulation (Eq. \ref{eq:PCformula}). \texttt{RV\_proxy} reports just the $R_V$ and its standard deviation according to the proxy formula from Section \ref{sec:rvdef}. Finally, \texttt{Qfactor} gives the quality factor $Q$, a masking fraction $M$ (Eq. \ref{eq:mfactor}), and an adjusted masking fraction 
\begin{equation}
M' = M M_E,
\label{eq:MME}
\end{equation}
where $M_E$ is roughly $S(E-0.2)$, a sigmoid function centered on the $E(B-V)$ threshold $0.2$. We use $M'$ as a value mask in the $R_V$ plots in Figures \ref{fig:RVoph}--\ref{fig:RVori}.

\begin{table*}
	\caption{$R_V$ to various clouds}
	\label{tab:clouds} 
  \begin{tabular*}{\textwidth}{ @{\extracolsep{\fill}} | l | r | r | c | c | }
	  \hline
    Target & locus-shift $R_V$ & literature $R_V$ & $(l,b)$ & reference  \\
		\hline
    Perseus & $3.3$--$4.1$   & $3$--$5$          & $(158 \pm 2, -20 \pm 3)$ & \cite{RVfost} \\
    Oph core & $4$--$6$      & $\chi^2(5.5) < \chi^2(3.1)$ $^\ast$ & $(353,18)$ & \cite{RVchap} \\
		Oph cloud & $3.6$--$4.4$ & $4.2\pm0.5$       & $(352\pm 3, 16 \pm 2)$ & \citet{Whittet1974} \\
		Alessi  95 & $~3.3$      & $2.80$ $\dagger$  & $(134,9)$   & \cite{TurnerAlessi} \\
		Pleiades   & $~3.5$      & $3.11$ $\dagger$  & $(167,-24)$ &\cite{TurnerP} \\
		NGC 1647   & $~3.5$      & $2.86$ $\dagger$  & $(180,-17)$ & \\
		Messier 4  & $~3.7$      & $3.76$ $\ddagger$ & $(351,16)$  & \cite{HendricksM4} \\
		Collinder 394 & $~2.5$   & $3.1$ $\dagger$   & $(15,-9)$   & \cite{TurnerCol} \\
    \hline  
  \end{tabular*}
	\tablecomments{Locus-shift R(V) estimates compared with selected results from the literature. \\ 
	$^\ast$\citet{RVchap} reports a lower $\chi^2$ for an $R_V$ of $5.5$ than for that of $3.1$ in the core of the Rho Ophiuchi molecular cloud. This roughly agrees with our result that pixels in the cloud have most likely $R_V$ values ranging from $4$ to $6$. \\
	$\dagger$ These estimates of $R_V$ rely on $UBV$ photometry whereas our PS1 estimates use $grizy$ photometry, and thus any comparison requires an extrapolation assuming some reddening law. \\
	$\ddagger$ \cite{HendricksM4} gives $R_V$ values determined by the formula $A(V)/E(B-V)$ as well as by the \citet{cardelli} law. We choose to show the former value here because we believe that parametrization is closer to ours.
\\}
\end{table*}

\subsection{$R_V$ for Nearby Dust Clouds}\label{sec:compare}

We compare our results with other studies that measure $R_V$ along a specific line of sight. In Table \ref{tab:clouds} we make our best effort to include all available papers as of this writing which (a) give results for a specific location or locations in the sky and (b) are covered by the PS1 footprint. There are several tens of papers such as \citet{GontRV} and \citet{SungRV} which provide spatial variations of $R_V$ but are not constrained enough to allow comparison with our results. After selecting for studies that overlap with the region of sky covered by PS1, we have a handful of targets. 
We find a broad range of estimated $R_V$ values from cloud to cloud, and although the numbers do not agree exactly, we find that the variation in our estimates are positively correlated with those from previous studies.

We also find some features of note in select clouds. E.g. we sometimes see a positive correlation between $R_V$ and $E(B-V)$ in clouds such as Ophiuchus (Figure \ref{fig:RVoph} ) but do not see them in areas like the clouds in Cepheus (Figure \ref{fig:RVcep}). If we had seen the former property consistently throughout the sky, it may have been due to artificial correlations in our model parameters. However, regions like Cepheus show very little correlation between $E(B-V)$ and $R_V$, and Perseus and Orion even exhibit the opposite correlation, albeit very subtly (Figures \ref{fig:RVper} and \ref{fig:RVori}). The different types of correlations show that our analysis is measuring a real reddening effect, and that it is potentially sensitive to complex properties of dust. The correlation in Ophiuchus itself may be due to its being at a stage of formation where its pressure and radiation field are conducive to large grain formation. We see a similar, but less pronounced, correlation between the dust column and $R_V$ in the densest cores of Taurus and Perseus. 

The dust in Cepheus also seems to consist of two distinct populations. Dust above $15$ degrees in galactic latitude $b$ has relatively low $R_V$, whereas the portion at $b < 15$ has elevated $R_V$. The separation along $b \sim 15$ corroborates the spatial structure of the Cepheus Flare described in \citet{SchlClouds} as well as \citet{CepGrenier} and \citet{CepKun}, providing further evidence that the Flare is two clouds along coincident lines of sight.

\section{The Reddening Law via Bayestar}\label{sec:bay}

Bayestar is a C++ package for performing Bayesian analysis on stellar photometry to determine the reddenings to and stellar types of a set of stars. It was developed by Gregory Green to produce a 3D map of dust in the Milky Way \citep{Green3D}. Please refer to \citet{greenMethods} and \citet{Green3D} for a complete overview of the techniques used by Bayestar. The following paragraphs describe a rather specific and limited application of Bayestar for the purpose of inferring $R_V$.

In the context of this study, we use Bayestar to test the efficacy of fitting for per-star stellar posteriors in order to estimate the reddening from dust along a line of sight. If one dust cloud along a line of sight that accounts for the majority of reddening, we expect the reddening estimate given by Bayestar to be close to that given by the locus-shift method. Since this use case is easy to understand, we attempt to show that Bayestar is a well-behaved and reliable estimator of reddening for such lines of sight. This should provide a foundation for future studies that will use better data and models. For example, when \textit{Gaia} data are released, we will be able to use distance constraints to more accurately estimate stellar type and reddening.

We note that in comparison to the locus-shift method, Bayestar assumes that the space of all possible reddenings (i.e. shifts in color space) is two-dimensional, and does not have the flexibility to explore arbitrary reddening laws. If all reddenings in the Milky Way do indeed lie in a two-dimensional space, then Bayestar naturally provides a better model for any reddening or selection effect that may arise owing to complex dust distributions and stellar populations along a line of sight. By comparing the two methods, we can corroborate one against the other, as well as determine the most efficient method for estimating reddening.

Our strategy with Bayestar is to fit for the stellar type (parameterized by $M_r$), metallicity $[Fe/H]$, distance modulus $\mu$, $E(B-V)$, and $R_V$ of a star given its magnitudes. The model developed in \citet{greenMethods} uses a function of $M_r$ and $[Fe/H]$ to get the intrinsic colors of a star and then adds the expected attenuation from $\mu$, $E(B-V)$, and $R_V$ to get the full model magnitude:
\begin{equation}
\vec{m}_{\textrm{mod}} = \vec{M}(M_r,[Fe/H]) + \vec{A}(E,R_V) + \mu.
\end{equation}
We have abbreviated $E(B-V)$ as the variable $E$ since the model does not treat it as a function of passbands, but rather as a parameter for specifying the dust column density to a star. $\vec{A}(E,R_V)$ is the reddening vector, which denotes the displacement in magnitude space due to dust extinction. The function $\vec{A}(E,R_V)$ itself is derived by integrating for each of the PS1 bands the expected reduction in flux from a typical main-sequence star due to dust:
\begin{equation}
A_b(E,R_V) = -2.5 \log\left[ \frac{\int{d\lambda W_b(\lambda)S(\lambda)f(\lambda,E,R_V)}}{\int{d\lambda W_b(\lambda)S(\lambda)}} \right]. \label{eq:redvec}
\end{equation}
$W_b$ is the band throughput provided by \citet{JTphoto}, $S$ is the spectrum for a typical star, and $f$ is the throughput of photons at $\lambda$ due to some quantity and type of dust specified by $E$ and $R_V$. That is, $f$ is determined by the extinction curve, which we define according to a reddening law such as F99. We then combine $A_b(E,R_V)$ for all bands $b$ to obtain $\vec{A}(E,R_V)$. A detailed treatment is available in the appendix of \citet{SchlaflyRed}. The likelihood is then
\begin{equation}
p(\vec{m}|M_r,[Fe/H],E,R_V,\mu) = N(\vec{m}|\vec{m}_{\textrm{mod}},\vec{\sigma}),
\end{equation}
where $N(\vec{m}|\vec{m}_{\textrm{mod}},\vec{\sigma})$ is a multivariate normal with mean $\vec{m}_{\textrm{mod}}$ and standard deviations $\vec{\sigma}$. Given some priors on the parameters, we can then obtain the posterior for a star. The posteriors for some stars with large uncertainties tend to either be multimodal or have an extremely elongated non-Gaussian shape, due to the surfaces mapped out by the locus model in parameter space. Please see the top panel of Figure \ref{fig:toyfit} for a graphical representation of the model and a visual explanation of how it differs from the locus-shift model.

We caution the reader that the model in this section assumes a specific reddening law. We use a principal component formulation that has been fit to F99 owing to evidence in favor of its being an accurate model for the PS1 bands \citep{SchlaflyRed}. However, this means that the reddening vectors probed by our Monte Carlo sampler only lie on a subsurface within the space of all possible extinctions. If some dust cloud were to have a reddening not well described by F99, then our fit for $R_V$ would not be reliable.

\subsection{Method for Estimating $R_V$ with Bayestar}\label{sec:baymethod}

Our method is as follows: We use a parallel affine-invariant sampler adapted from \citet{GoodmanWeare} and run chains for each star using 
the model described above \citep{greenMethods}. Stars are selected from the full list of sources based on whether there is detection in at least four bands and whether the point-spread function is point-like (i.e. it is not a galaxy). We use a kernel density estimator to find the marginal posterior as a function of $R_V$. A grid spacing of $0.05$ $R_V$ and a Gaussian kernel with FWHM $1.5$ times the grid spacing, i.e. $\sigma_{R_V} = .18$, give a resolution equal to the typical uncertainty in $R_V$ for a dust column of approximately $E(B-V) = 1$.

We then take the product of the $R_V$ posterior distributions of all the stars in a single pixel. Assuming that there must only be a single $R_V$ value for a sufficiently small pixel, this is the joint probability distribution of $R_V$ for all stars in a pixel. In reality we found that a few percent of the sources in a pixel tend to be outliers owing either to bad photometry or to not being a main-sequence star. In order to account for this, we modify each stellar posterior distribution to be a sum of the original marginalized distribution and an additional flat distribution from $R_V = 1$ to $R_V = 9$ that has been normalized to have a total integrated probability $.05$ times that of the marginalized distribution, i.e., $P'(R_V) \propto P(R_V) + .00625\int{P(R_V)}$.

The end result is an $R_V$ distribution for every pixel in the sky that contains PS1 sources. We note that since we have marginalized over the distance modulus, this map does not have 3D information like the ones published by the related work \citet{greenMethods}. This is an intentional simplification for the purposes of drastically decreasing computational time and for making a more direct comparison with the alternate reddening maps in this paper, which are both 2D.

Our distribution of $R_V$ values agrees with most publications in the literature. It is centered around $3.3$ with an FWHM of $0.5$. There is also no significant correlation with the other fit parameters, except for a slight increase for very high values of $E(B-V)$, which could be due to the dust grain population. We compare our Bayestar results with our locus-shift method in Section \ref{sec:analyze}. We note that in order to make such a comparison, we require some method for estimating a degree of belief for the reddenings we find on a pixel-by-pixel basis. We use the evidence for this purpose. We provide the details of our method for calculating it in the Appendix.

\section{Discussion}\label{sec:analyze}

\subsection{The Reddening Law}\label{sec:redlaw}

\begin{figure}
   \includegraphics[height=3.6in,angle=90]{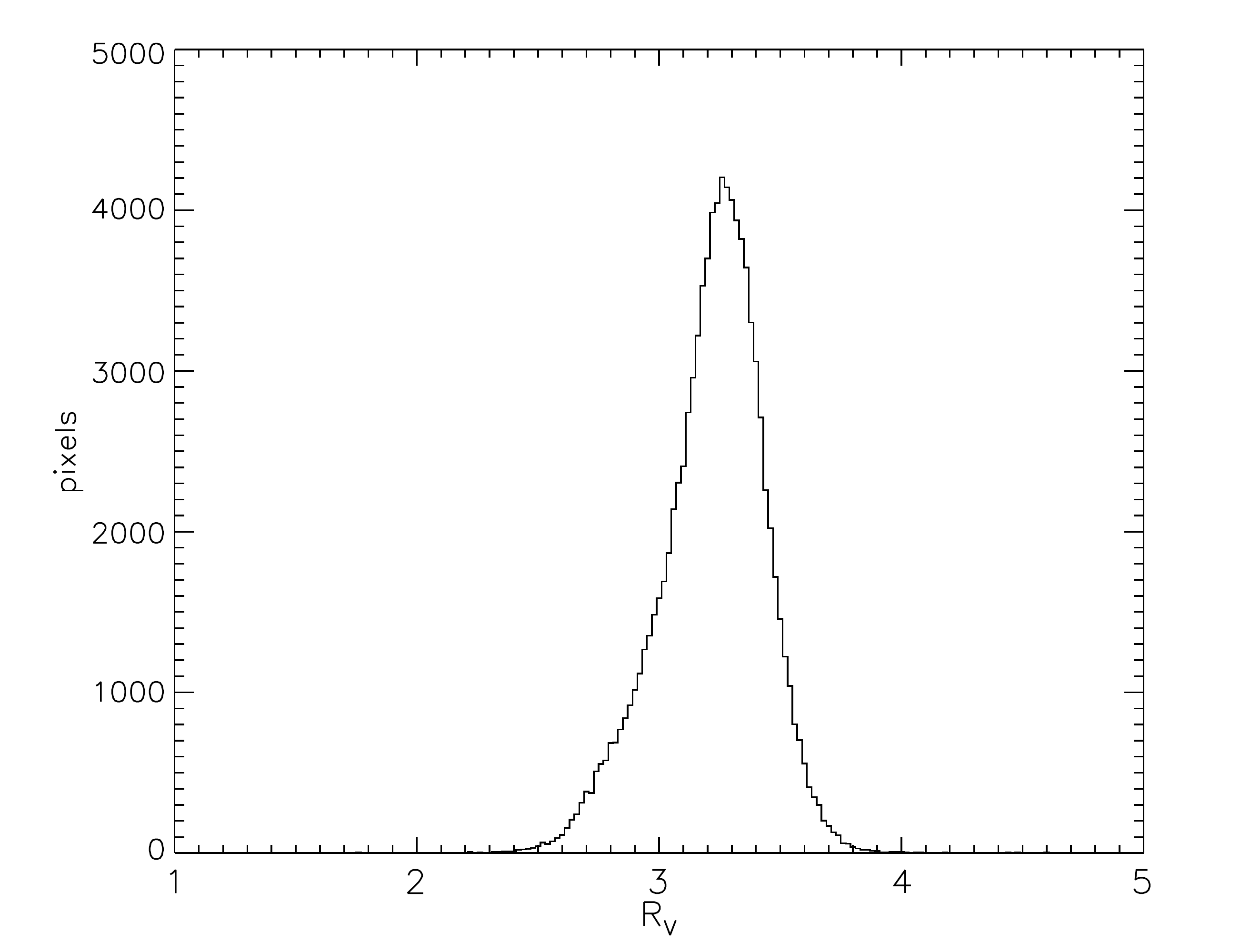} 
   \caption{$R_V$ distribution for all pixels shown in the locus-shift map. We find a mean $R_V$ of $3.28$ and an FWHM of $0.5$. The distribution has similar properties to those reported in \citetalias{apogee}, albeit with a larger wing for $R_V < 3$. However, we note that due to the dependence on the definition of the reddening law one uses to convert colors to $R_V$, there is up to a factor of $2$ in uncertainty, mostly from a linear scaling.  \\}
\label{fig:RVdist}
\end{figure}

\begin{figure}
	 \includegraphics[height=3.6in,angle=90]{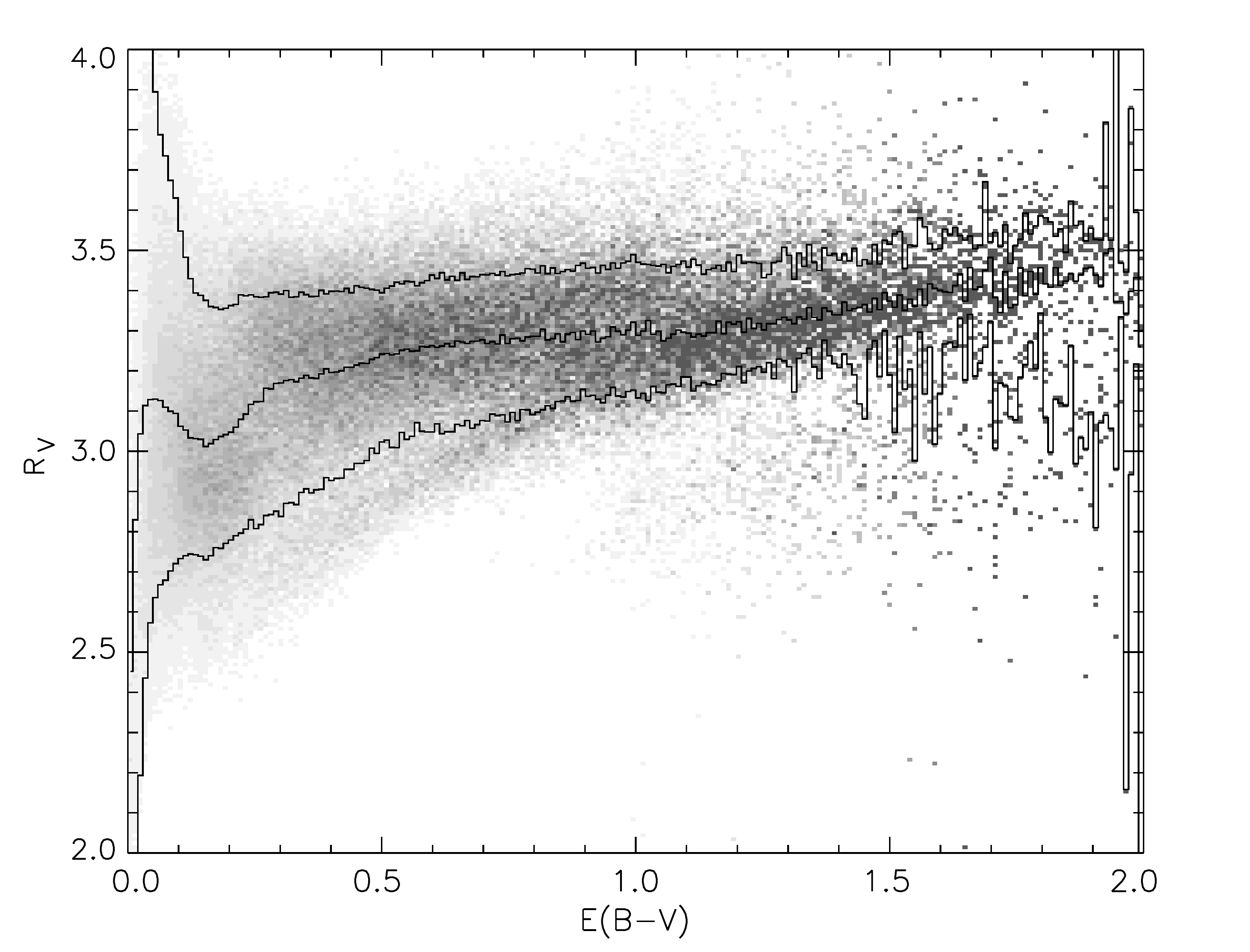} 
   \caption{$R_V$ vs. $E(B-V)$ for the locus-shift method. On the whole, there is no significant correlation between $R_V$ and $E(B-V)$ throughout the mid-Galatic latitudes of the Milky Way, except for a slight average increase in $R_V$ for larger values of $E(B-V)$. Figures \ref{fig:RVoph}--\ref{fig:RVori} show that this is not indicative of any problematic dependencies in the parameters of our reddening model. However, some of the variation in $R_V$ may be correlated with proximity to the Galactic center, as detailed in Section \ref{sec:redlaw}.\\ \\}
\label{fig:RVvE_dist}
\end{figure}

Among the more widely cited reddening laws, we find F99 to be the best fit to our locus-shift results. Converting our locus shifts to $R_V$ via the proxy in Equation \ref{eq:gygr}, we find the mean $R_V$ to be $3.28$ with an FWHM of $0.5$ (Figure \ref{fig:RVdist}). This is in agreement with the distribution found by S16. As described in Section \ref{sec:choose}, we fit F99 to the locus shifts via gradient descent in order to obtain the corresponding $E(B-V)$ values. We find that there are three populations of pixels roughly divided at $E(B-V)$ values of $0.23$ and $0.75$ and lying above or below the median $R_V$ line (Figure \ref{fig:RVvE_dist}). It is not yet clear at the present level of analysis whether this is a by-product of having an incomplete model or the result of varying properties in different populations of dust. Pixels lying above the median line tend to be closer to the direction of the Galactic anticenter, whereas pixels in the two populations below the line tend to be closer to the Galactic center and exhibit a bimodal distribution in $R_V$ and $E(B-V)$ space. In the current data product there is too much spatial mixing for us to conclude that this is a real physical feature. 

In general, $R_V$ increases slightly for larger column densities but is mostly independent of $E(B-V)$, as demonstrated by our detailed examination of Ophiuchus and Cepheus, as well as the distribution shown in Figure \ref{fig:RVvE_dist}. However, there are localized correlations, and in particular, some dense cloud cores exhibit highly elevated $R_V$ values.
This effect exceeds any variations in our mock catalogs (see Section \ref{sec:mock}), and we have reason to believe that it is real since other studies have found similar properties in dust clouds. A frequently cited reason for this correlation is that the higher concentrations of dust grains in dense clouds may facilitate the formation of larger grains, which in turn shield against UV radiation that may destroy dust mantles. The larger grain population results in a shallower extinction curve in the visible wavelengths owing to Rayleigh and Mie scattering. 

We also note that, although mostly masked out by our recommended quality factor threshold, pixels near the Galactic plane tend to have elevated $R_V$, except for those near the bulge, which have low $R_V$. These are regions where our model is not a good descriptor of the data, due to both varying stellar populations and multiple layers of dust. In fact, the systematic offset may be caused by our fits being sensitive to such differences in the properties of the stars and dust. This may suggest an opportunity to extract even more information from PS1 photometry given more sophisticated models.

\begin{figure}
   \includegraphics[width=3.8in]{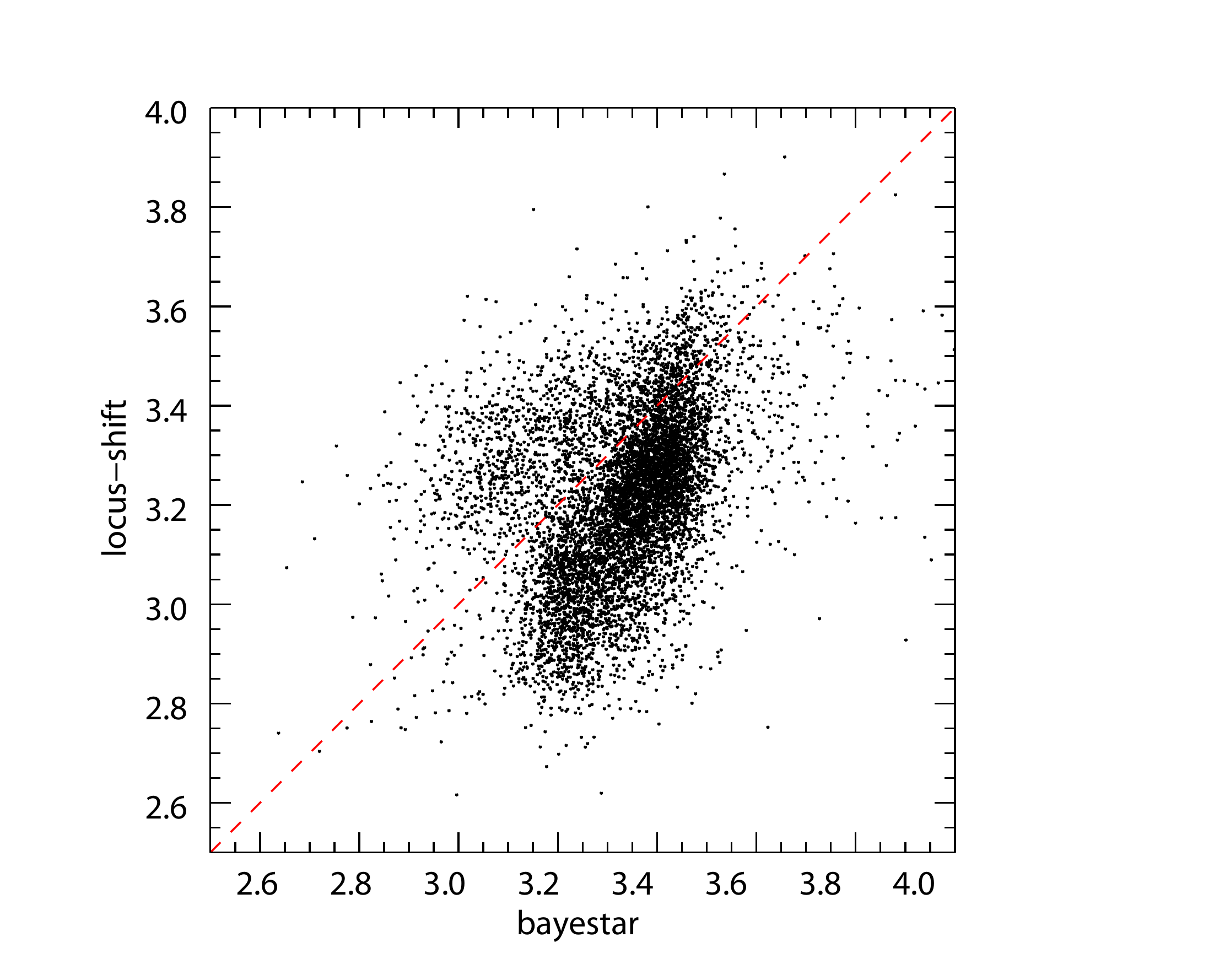} 
   \caption{$R_V$ analog from the locus-shift method vs. $R_V$ from Bayestar. Each point shows the estimated $R_V$ for a Healpix pixel using the locus-shift and Bayestar algorithms.  The dashed red line is provided as reference for an identity relation. We selected clouds at high Galactic latitudes that passed the quality cuts from the locus-shift run to make this comparison. Although the two methods use fundamentally different projections of the stellar band magnitudes, there is a clear correlation between the two methods over the whole sky. That is, estimates of reddening are robust even when magnitudes are combined differently. We note that the Bayestar $R_V$ values, which were originally estimated using the \citetalias{apogee} principal components, have been appropriately converted to match the parameterization used in the locus-shift method. Therefore, the slope in the relation is not due to the effect described in Figure \ref{fig:redlaw}, but rather a result of using different models and necessarily different priors, the details of which await further analysis.\\}
\label{fig:ShiftvBay}
\end{figure}

\subsection{Bayestar versus Locus Shift}\label{sec:bayLS}

We find agreement between our results for the Bayestar method and the locus-shift method. In Figure \ref{fig:ShiftvBay} we plot $R_V$ values from Bayestar against those from the locus-shift method.  Each point is an $R_V$ comparison for a Healpix pixel. Although individual clouds do not show a strong correlation, the coarse spatial variation of $R_V$ across the whole sky is consistent from cloud to cloud. Since the two methods are largely independent, we have some confidence that we are estimating real variations in the reddening law. 

In order to make this comparison, we have to convert from locus shifts to a definition of $R_V$ close to what is modeled by Bayestar's reddening law. In order to facilitate this, we parameterize Bayestar's allowed reddenings according to the space mapped by the first two principal reddening components from \citetalias{apogee}. This covers a subspace in color space that is very similar to that mapped by F99 but has the additional advantage of being linear. This in turn allows us to convert locus shifts into corresponding $R_V$ values using simple linear transformations, as described in Section \ref{sec:linear}.

The above procedure makes our two derivations of $R_V$ consistent. However, the output of the locus-shift method after the conversion to $R_V$ is a single chain of $R_V$ values for each pixel. On the other hand, the output of our Bayestar method is a chain of $R_V$ values for every single star in each pixel. Therefore, we take the product of the $R_V$ distributions of all the stars in each pixel (see Section \ref{sec:baymethod}).

The main discrepancies between the two methods in Figure \ref{fig:ShiftvBay}, i.e. the slope and the offset in the mean $R_V$ values, can probably be attributed to the fact that - despite our best efforts as described above - it is impossible to equate the two results since the respective Monte Carlo chains project the probabilities associated with a pixel into different subspaces. To be more precise, Bayestar projects reddening information into the subspace of extinctions mapped by F99 \textit{first} before fitting for individual stellar likelihoods and subsequently calculating the likelihood of all stars in a pixel, whereas the locus-shift method evaluates the combined likelihood of all the stars in a pixel while fitting for the full reddening distribution in color space. Therefore, Bayestar is losing reddening information, while the locus-shift method is losing information from individual stars. Furthermore, the slope of the relation is fairly sensitive to our choice of priors for either method, and although we keep all shared priors identical for our analysis, there are unique priors for which we must ultimately provide a best guess based on other literature (e.g. priors for metallicity or the full reddening vector). With this context, it is perhaps more assuring that we have a significantly positive correlation at all.

\subsection{PS1 versus APOGEE}
	
\begin{figure}	
   \includegraphics[width=3.2in]{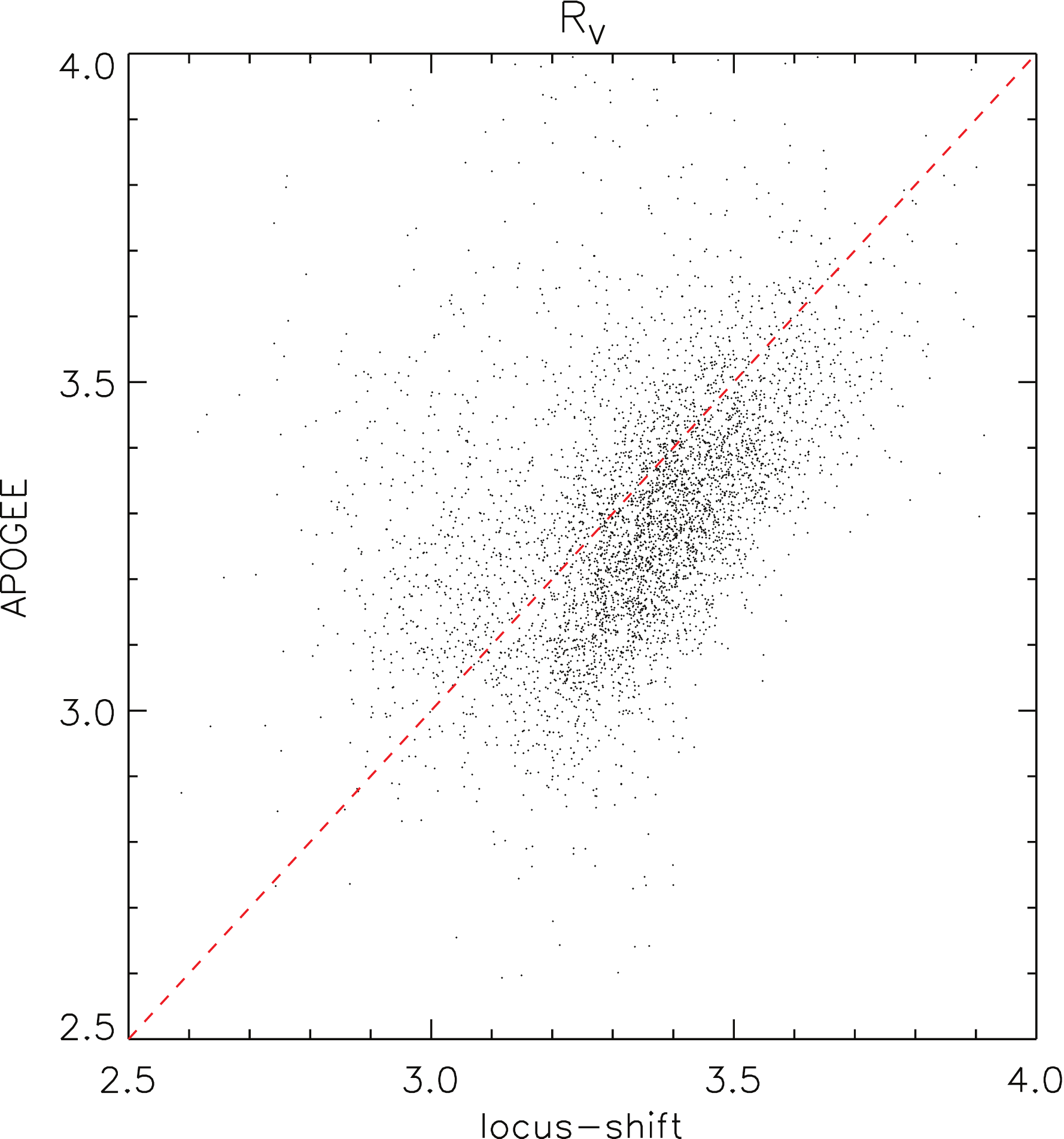}  
	 \vspace{.5cm}\\
   \includegraphics[width=3.2in]{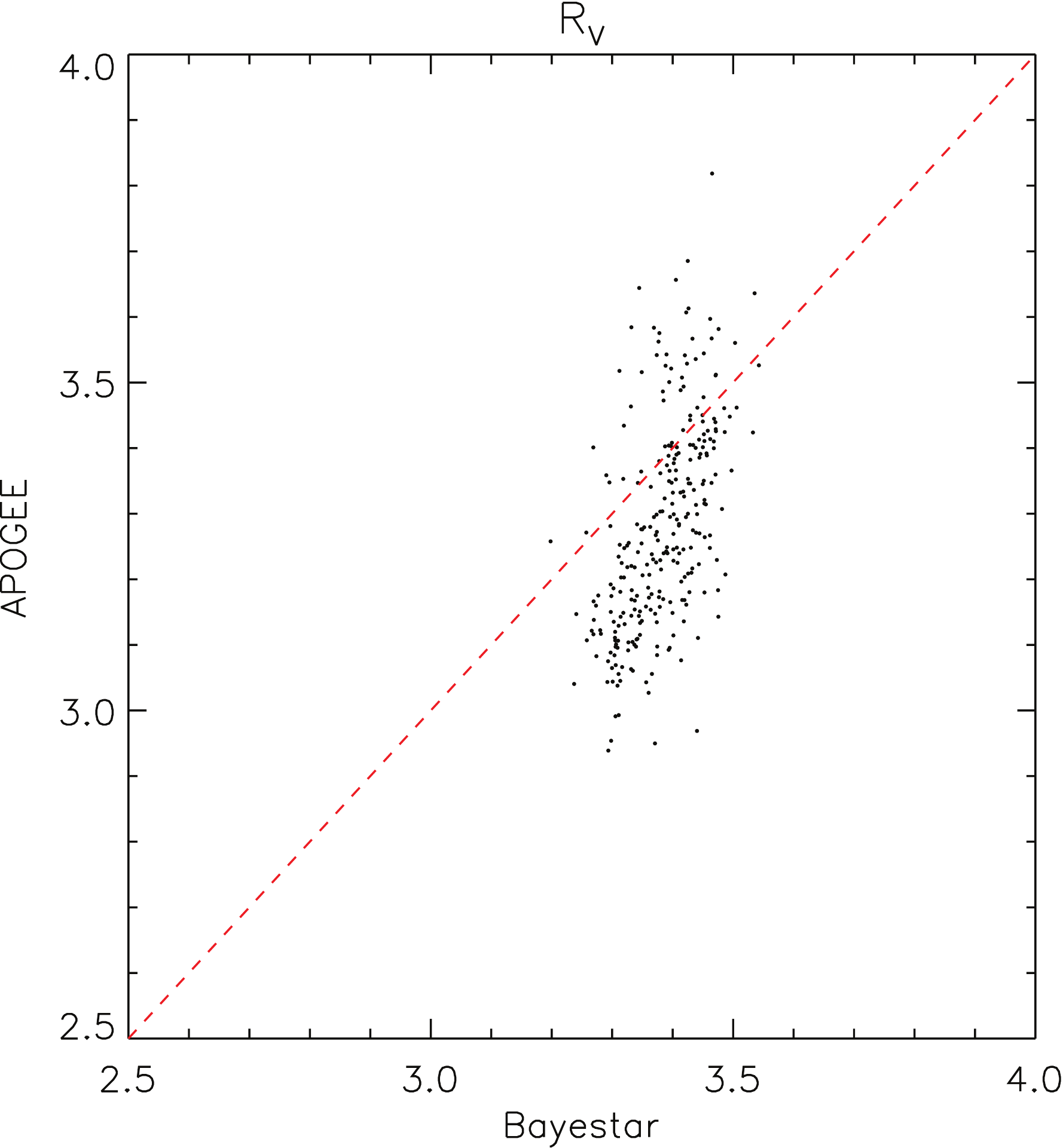} 
   \caption{In the top panel we compare $R_V$ values inferred from the locus-shift method with those provided by \citetalias{apogee}. In the bottom panel we do the same with values inferred using the Bayestar method. Since the locus-shift and Bayestar methods have different degrees of confidence for individual pixels, we use separate quality cuts when making the comparison to the APOGEE-based results. We note that all $R_V$ values in the APOGEE-locus-shift comparison were calculated only using the $g$, $r$ and $y$ bands. This means that the comparison should not be affected by the type of problematic parameterizations shown in Figure \ref{fig:redlaw}. For the Bayestar comparison, however, we had to fit to the data presupposing F99, which means that the range of possible reddening vectors explored by the sampler was more constrained, as evident in the scatter of values. The fact that neither plot shows a slope of $1$ is probably a result of the studies necessarily using different sets of stars. Nevertheless, we see a positive correlation between all three results, which supports the hypothesis that we are sensitive to real reddening information within the stellar spectra.\\}
\label{fig:apoVS}
\end{figure}

\begin{figure}
   \includegraphics[height=3.6in,angle=90]{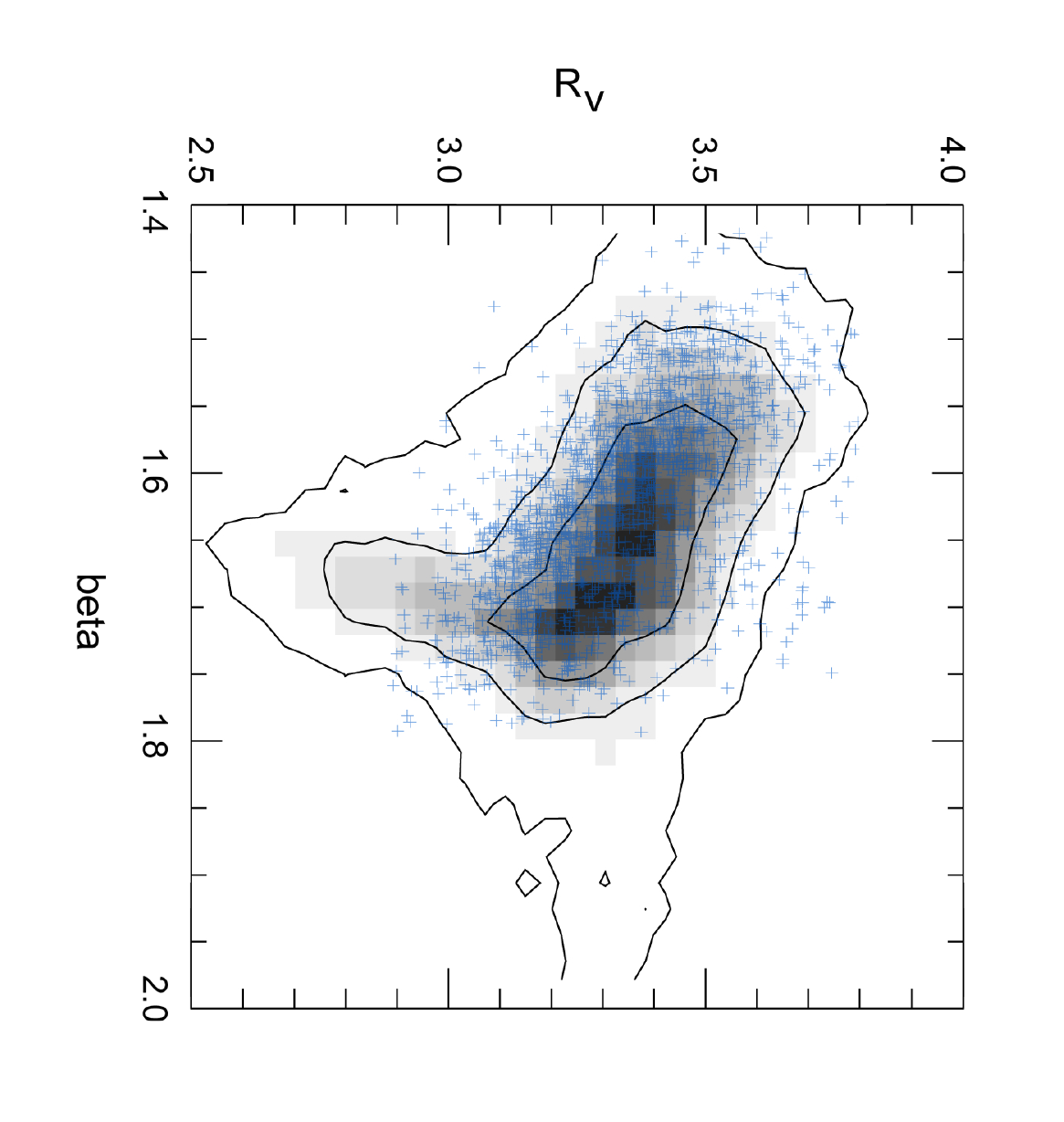}
	 \caption{$R_V$--$\beta$ correlation between locus-shift results and the Planck collaboration's estimate of $\beta$. We display the data as a 2D histogram of $R_V$ and $\beta$ values from the pixels shown in the bottom panel of Figure \ref{fig:f7}, with darker bins denoting higher densities of points. We also show lines approximating isocontours of the point density. The histogram is very similar to the distribution of points in Figure 18 of \citetalias{apogee}, which we have directly copied from \citetalias{apogee} and overlaid as blue plus signs. This gives us confidence that both studies are measuring real physical properties of dust. \\}
\label{fig:RVbeta}
\end{figure}

\begin{figure}
   \includegraphics[width=3.5in]{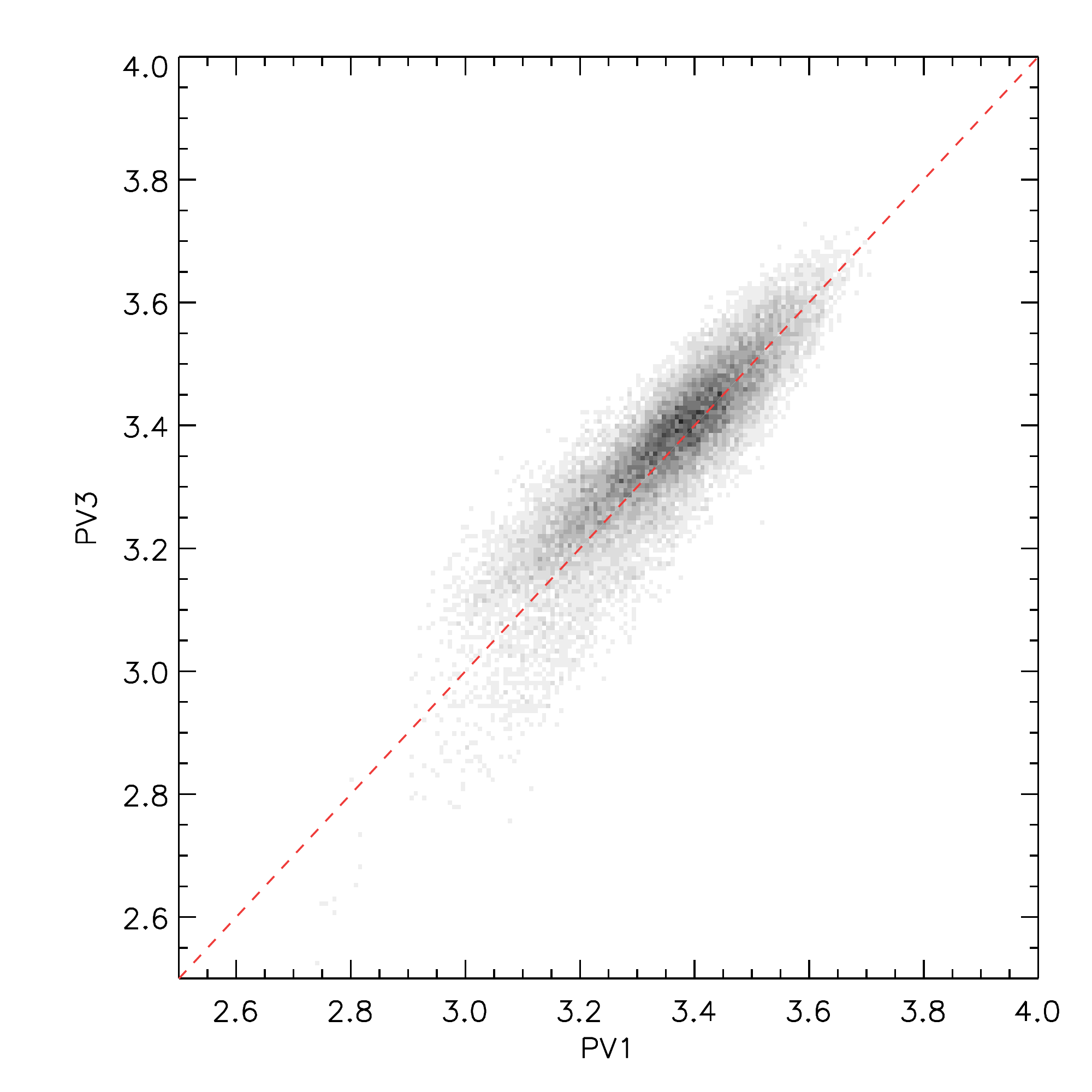}
   \caption{Locus-shift $R_V$ values for Healpix pixels using PS1 PV3 data compared to those using PV1 data. The spread of values here shows how sensitive our fits are to even a couple hundredths of magnitude's difference between data releases.\\}
\label{fig:PV3PV1}
\end{figure}

We expect the reddening maps from \citetalias{apogee} to be reliable not only because they are empirically determined from photometry, like our locus-shift map, but also because the stellar types of the sources were fit independently of the dust reddenings to the sources. This was made possible by the detailed temperature information made available by the APOGEE survey. By tying the stellar type fits to the temperature, one can decouple reddening from stellar types and obtain better constraints on $R_V$. In this regard we expect comparisons with \citetalias{apogee} to be the strongest tests currently available of the reliability of the locus-shift and Bayestar methods. This of course will change when more \textit{Gaia} data are released.

We find $R_V$ to be correlated for the different surveys and methods. We find a positive correlation between the results from \citetalias{apogee} and the $R_V$ values from both our Bayestar and locus-shift methods	 (Figure \ref{fig:apoVS}). Although the slope between the Bayestar and S16 results shows a correlation not as strong as that between locus-shift and S16, both comparisons clearly demonstrate that all three surveys are sensitive to the same reddening information. 

We also corroborate the negative correlation between $R_V$ and Planck $\beta$ that was discovered by S16. In Figure \ref{fig:RVbeta} we show the $R_V$--$\beta$ relationship from our results. The location and shape of the distribution are strikingly similar to those in Figure 18 of S16. We also find that this relationship persists independently of cuts on latitude or dust column.

We additionally find $R_V$ to be consistent between different PS1 processing versions (Figure \ref{fig:PV3PV1}), which shows that our estimates of reddening are robust even under different reductions of the PS1 surveys.

\section{Conclusion}\label{sec:conclude}

Historically it has been difficult to measure reddenings reliably over the entire sky owing to the lack of a consistent data set with sufficient coverage. We present one of the first maps of reddening variation in the Milky Way for a majority of the sky. This is the first reddening map based purely on PS1 photometry, as well as the first to cover more than half of the sky at a $15$-arcminute resolution. Comparisons of our locus-shift results with other reddening measurements show that careful modeling of the photometry of typical stars allows us to obtain mutually consistent estimates of the spatial variation in dust reddening. We publish the map as a set of Healpix pixels, providing both the full reddening vectors (i.e. locus shifts) and the converted $R_V$ values. 

Upon inspection of the maps, we observe the following:

\begin{enumerate}
\item The Bayestar and locus-shift methods obtain estimates of $R_V$ that are reasonably correlated, despite using very different models and algorithms.
\item Mock data show the locus-shift method to be self-consistent in the limits where it is expected to agree with Bayestar.
\item We find the $R_V$ distribution and spatial variation in the reddening maps to be consistent with \citet{fitz} and \citet{apogee}, respectively.
\item We find further agreement with several detailed studies of $R_V$ in a set of well-characterized nearby dust clouds.
\end{enumerate}

Given the level of self-consistency, as well as agreement with other studies, we believe that our algorithms are probing the actual reddenings of stars. We have demonstrated that Bayestar and the locus-shift method reliably estimate $R_V$ in all the limiting cases we can easily check. In particular, with the right models and data, even a simple locus-regression algorithm can be an effective tool for quickly estimating the reddening law. We expect the locus-shift method to be a useful sanity check when making a full 3D map of reddening with Bayestar, and we aim to extend Bayestar so that it can be sensitive to all dimensions of reddening along a line of sight. We should be able to corroborate future results from Bayestar with 3D studies like \citet{schlafly3D}, as well as surveys with better distance information such as \textit{Gaia}. Altogether, we have shown that it is possible to use large photometric surveys to make multiple independent estimates of reddening across the Milky Way.

\subsection{Acknowledgments}

A.L. acknowledges support for this research provided by the NSF. This material is based on work supported by the National Science Foundation Graduate Research Fellowship under grant no. DGE 1144152. This work is also supported by NSF grants AST-1312891 and AST-1614941. E.S. acknowledges support for this work provided by NASA through Hubble Fellowship grant HST-HF2-51367.001-A awarded by the Space Telescope Science Institute, which is operated by the Association of Universities for Research in Astronomy, Inc., for NASA, under contract NAS 5-26555.

We would like to thank the Pan-STARRS1 collaboration for providing us with high-quality photometry for millions of stars across $3/4$ of the sky. The Pan-STARRS1 Surveys (PS1) and the PS1 public science archive have been made possible through contributions by the Institute for Astronomy, the University of Hawaii, the Pan-STARRS Project Office, the Max-Planck Society and its participating institutes, the Max Planck Institute for Astronomy, Heidelberg, and the Max Planck Institute for Extraterrestrial Physics, Garching, The Johns Hopkins University, Durham University, the University of Edinburgh, the Queen's University Belfast, the Harvard-Smithsonian Center for Astrophysics, the Las Cumbres Observatory Global Telescope Network Incorporated, the National Central University of Taiwan, the Space Telescope Science Institute, the National Aeronautics and Space Administration under grant no. NNX08AR22G issued through the Planetary Science Division of the NASA Science Mission Directorate, the National Science Foundation grant no. AST-1238877, the University of Maryland, Eotvos Lorand University (ELTE), the Los Alamos National Laboratory, and the Gordon and Betty Moore Foundation.

The analyses presented in this paper were run on the Odyssey cluster at Harvard University. We would like to acknowledge the staff at the FAS Research Computing Group, who not only manage the Odyssey cluster but also provide maintenance on our own dedicated machines.

We thank Mario Juric for building the Large Survey Database framework, which powers all the queries made to our local copy of the PS1 data set. 

This research has made use of NASA's Astrophysics Data System Bibliographic Services.

Finally, we appreciate the many useful comments from Aaron Meisner, Stephen Portillo, Tansu Daylan, Catherine Zucker, Ioana Zelko, Blakesley Burkhart, Zachary Slepian, Ana Bonaca, Josh Speagle, Ben Lee, Karin Oberg, and Vinothan Manoharan. 

\bibliography{RVrefsArxiv}

\begin{appendices}
\section{Harmonic Mean Estimation of the Evidence}\label{appendix:evidence}

The harmonic mean estimator allows us to calculate the evidence of some data by only using the likelihoods and priors of samples in a chain. Starting with a simple relation for the Bayesian evidence $Z(D)$, we can derive the expression as follows:

\begin{eqnarray}
P(\theta|D)Z(D) &=& L(D|\theta)\Pi(\theta), \\
\frac{1}{Z(D)} &=& \frac{P(\theta|D)}{L(D|\theta)\Pi(\theta)}, \\
&=& \frac{P(\theta|D)}{L(D|\theta)\Pi(\theta)} \int{d\theta' \phi(\theta')}
	\label{eq:phi_int}, \\
&=& \int{ d\theta' \frac{P(\theta'|D)}{L(D|\theta')\Pi(\theta')} \phi(\theta')}, \\
&=& \int{ d\theta' \frac{\phi(\theta')}{L(D|\theta')\Pi(\theta')} P(\theta'|D)}, \\
&=& \frac{1}{N} \sum_{ \theta_i \in \textrm{\small{ chain}} } \frac{\phi(\theta')}{L(D|\theta')\Pi(\theta')}.
\label{eq:evidence}
\end{eqnarray}

In Equation \ref{eq:phi_int}, the regulating function $\phi$ is an arbitrary normalized function. If the factor $P(\theta|D) / L(D|\theta)\Pi(\theta)$ is well behaved over the entire domain of $\phi$, then we can bring it inside the integral since by definition it should be a constant. After some rearranging, we get an expectation of samples drawn from the posterior $P(\theta'|D)$. In practice, we can use the relation \ref{eq:evidence} to estimate the evidence as long as the regulating function falls to zero faster than the posterior \citep{2009AIPC.1193..251R}.

Because the distribution of the posterior in $(E(B-V),\mu)$ space is highly irregular, we must be careful about picking a valid regulating function. We opt to use a constant four-dimensional ellipsoid that is zero outside its boundaries and centered at the highest likelihood sample from the Monte Carlo chains. This ensures that there are a high number of samples in that region of parameter space and that the posterior will be well behaved in the immediate vicinity. Making the boundary of $\phi$ be a step function ensures that it falls off faster than the posterior, and it makes it easy to normalize. Since the shape of the posterior can vary drastically in different parts of parameter space, we recalculate the local covariance matrix of samples near the regulating region so that we can update the principal axes of the ellipsoid to better approximate the posterior.

Since we do not let the Monte Carlo chains have negative reddening, the prior is effectively $0$ for $E(B-V) < 0$. Thus, we must make sure not to include this region in our regulating function. We achieve this by requiring the ellipsoid center to be at least some distance from $E(B-V)=0$, and if necessary, we renormalize the regulating function after ignoring the fraction of its volume in negative $E(B-V)$.

The accuracy of the harmonic estimator is primarily limited by Poisson statistics since we are effectively counting the number of samples that fall inside the regulating ellipsoid. This requires us to run longer chains since generally only $5-20\%$ of samples fall inside the ellipsoid. Of course, in pixels with a large number of stars we get a $\sqrt{N}$ reduction in this noise when determining the total probability of the pixel:

\begin{equation}
Z_{pixel} = \prod_j{Z(D_j)}.
\end{equation}

We find $10000$ steps to be sufficient for most pixels. To further reduce the noise, we use multiple regulating ellipsoids for each chain. We choose the location of each ellipsoid iteratively: we pick the highest-likelihood sample as the center of the first ellipsoid, then we pick the next-highest-likelihood sample that is outside some small exclusion zone around the first ellipsoid, etc. This ensures that the ellipsoids will all be different and all in regions with high sample density.

\end{appendices}

\end{document}